\documentclass[10pt,twocolumn,twoside]{IEEEtran}
\IEEEoverridecommandlockouts
\usepackage{cite}
\usepackage{algorithmic}
\usepackage{textcomp}
\usepackage{xcolor}
\usepackage{soul}
\usepackage{graphicx,color}
\usepackage{amsfonts,amsthm,amssymb}
\usepackage{mathrsfs,amsmath,arydshln,latexsym}
\usepackage{autobreak}
\usepackage{cite,url}
\usepackage[bookmarks,colorlinks]{hyperref}
\usepackage[linesnumbered,ruled,vlined]{algorithm2e}
\usepackage{multirow,array,makecell}
\usepackage{stfloats}
\usepackage{fancyhdr}
\usepackage{bm}
\usepackage{enumitem}
\usepackage{pifont}
\usepackage{subcaption}
\usepackage{nomencl}
\usepackage{colortbl}
\usepackage{diagbox}
\usepackage{balance}
\usepackage{braket}
\usepackage{physics}
\usepackage{framed}
\usepackage{threeparttablex}

\allowdisplaybreaks

\newcommand{\tabincell}[2]{\begin{tabular}{@{}#1@{}}#2\end{tabular}}

\begin{document}

	\title{Rydberg Atomic Quantum Receivers for\\ Classical Wireless Communications and Sensing:\\ Their Models and Performance
	}

	\author{Tierui Gong,~\IEEEmembership{Member,~IEEE}, 
		Jiaming Sun, 
		Chau Yuen,~\IEEEmembership{Fellow,~IEEE}, 
		Guangwei Hu, 
		Yufei Zhao, \\
		Yong Liang Guan,~\IEEEmembership{Senior Member,~IEEE},
		Chong Meng Samson See,~\IEEEmembership{Member,~IEEE},\\
		Mérouane Debbah,~\IEEEmembership{Fellow,~IEEE},
		Lajos Hanzo,~\IEEEmembership{Life Fellow,~IEEE}
		\vspace{-1cm}
		\thanks{T. Gong, J. Sun, C. Yuen, G. Hu, Y. Zhao, and Y. L. Guan are with School of Electrical and Electronics Engineering, Nanyang Technological University, Singapore 639798 (e-mail: trgTerry1113@gmail.com, n2308800l@e.ntu.edu.sg, chau.yuen@ntu.edu.sg, guangwei.hu@ntu.edu.sg, yufei.zhao@ntu.edu.sg, eylguan@ntu.edu.sg). 
		C. M. S. See is with DSO National Laboratories, Singapore 118225 (e-mail: schongme@dso.org.sg).
		M. Debbah is with the Center for 6G Technology, Khalifa University of Science and Technology, Abu Dhabi, United Arab Emirates (e-mail: merouane.debbah@ku.ac.ae). 
		L. Hanzo is with School of Electronics and Computer Science, University of Southampton, SO17 1BJ Southampton, U.K. (e-mail: lh@ecs.soton.ac.uk).
		}
	}
	
	\maketitle

	\begin{abstract}
		The significant progress of quantum sensing technologies offer numerous radical solutions for measuring a multitude of physical quantities at an unprecedented precision. Among them, Rydberg atomic quantum receivers (RAQRs) emerge as an eminent solution for detecting the electric field of radio frequency (RF) signals, exhibiting great potential in assisting classical wireless communications and sensing. So far, most experimental studies have aimed for the proof of physical concepts to reveal its promise, while the practical signal model of RAQR-aided wireless communications and sensing remained under-explored. Furthermore, the performance of RAQR-based wireless receivers and their advantages over classical RF receivers have not been fully characterized. To fill these gaps, we introduce the RAQR to the wireless community by presenting an end-to-end reception scheme. We then develop a corresponding equivalent baseband signal model relying on a realistic reception flow. Our scheme and model provide explicit design guidance to RAQR-aided wireless systems. We next study the performance of RAQR-aided wireless systems based on our model, and compare them to classical RF receivers. The results show that Doppler broadening-free RAQRs are capable of achieving a substantial received signal-to-noise ratio (SNR) gain of over $27$ decibel (dB) and $40$ dB in the photon shot limit and standard quantum limit regimes, respectively. 
	\end{abstract}

	\begin{IEEEkeywords}
		Rydberg atomic quantum receiver (RAQR), wireless communication and sensing, equivalent baseband signal model, photon shot limit (PSL), standard quantum limit (SQL)
	\end{IEEEkeywords}

	\vspace{-1em}
	\section{Introduction}
	
	The growing thirst for high data rate, large-scale connectivity, and ultra-reliable low latency are the distinctive features of the next-generation (NG) wireless systems, in which a multitude of upper-layer applications and a wide variety of functionalities over a large spatial distribution are expected to be supported \cite{recommendation2023framework}. To support these visions, developing advanced high-sensitivity receivers and exploiting a wide range of untapped spectral resources becomes paramount. Both require the radio-frequency (RF) receivers capable of detecting RF signals with extremely low strength and broadband spectrum. To realize these visions, the conventional technology road-map relies on the advances of integrated circuits (ICs) and antenna technologies.

	State-of-the-art RF receivers rely on well-calibrated antennas, filters, amplifiers, and mixers \cite{Moghaddasi2020Multifunction}. The sensitivity of IC-based receivers is typically limited by the extrinsic noise background, the bandwidth, the noise figure of ICs, and the signal-to-noise ratio (SNR) required for demodulation.  Therefore, designing high-quality ICs becomes an effective technique of increasing the sensitivity of RF receivers. Furthermore, strong reception capability may be achieved by harnessing multiple antennas, where multiple copies of an RF signal can be obtained for suppressing noise contamination and/or wireless channel fading. The emergence of multiple-input multiple-output (MIMO) systems relies on sophisticated multiple-antenna technology. Their concept has evolved from the massive MIMO philosophy \cite{Larsson2014Massive} to the development of holographic MIMO \cite{gong2023holographic, Gong2024Near}. For broadband signal reception and processing, the RF receivers may be implemented by relying on multiple IC channels, where each channel is responsible for a specific frequency band. Such a naive IC stacking scheme is bulky, power-thirsty, and heavily relies on advanced IC manufacturing technologies, suffering from many design challenges, particularly when designing complex high-frequency receivers.

	The ``second quantum revolution" \cite{dowling2003quantum} rests on three main pillars of quantum technologies, which are quantum computing, quantum communications, and quantum sensing. Specifically, quantum sensing includes but it is not limited to quantum precision measurements and quantum metrology, which rely on the use of quantum phenomena to perform a measurement of a physical quantity \cite{degen2017quantum}. The physical quantity is associated with the intrinsic properties of microscopic particles that are capable of facilitating an unprecedented level of sensitivity. Based on this, a multitude of quantum sensing applications have emerged for measuring a wide variety of physical quantities, such as electric fields, magnetic fields, frequency, temperature, pressure, rotation, acceleration, and so forth \cite{gschwendtner2024quantum}. Among the various quantum sensing technologies, those relying on Rydberg atoms exhibit an exceptional sensitivity in detecting electric fields \cite{schlossberger2024rydberg, zhang2024rydberg, Fancher2021Rydberg}, paving the way for realizing Rydberg atomic quantum receiver (RAQR) aided wireless communications and sensing \cite{gong2024RAQRs}. Additionally, the ambition of using quantum-domain solutions for enhancing the performance of classical wireless systems was documented for example in \cite{Botsinis2013Quantum,Botsinis2019Quantum}.

	Apart from their extremely high sensitivity, RAQRs have numerous compelling advantages, such as broadband tunability, compact form factor, and International System of Units (SI)-traceability \cite{Fancher2021Rydberg, gong2024RAQRs}. Hence they offer a radical solution to those RF reception challenges. Firstly, the high sensitivity of RAQRs is a direct benefit of the highly-excited state of Rydberg atoms, where their outermost electrons are excited to a very high energy level. The highly-excited states exhibit a remarkably large dipole moment that is extremely sensitive to the external coupling of RF signals. The high sensitivity of RAQRs has been experimentally shown to be on the order of \textmu V/cm/$\sqrt{\text{Hz}}$ using a standard structure \cite{sedlacek2012microwave, fan2015atom, Fancher2021Rydberg}. This has also been improved to $55$ nV/cm/$\sqrt{\text{Hz}}$ using a superheterodyne structure \cite{simons2019rydberg, jing2020atomic}, and can be further enhanced via repumping \cite{prajapati2021enhancement}. Another recent work further enhanced the sensitivity to $3.98$ nV/cm/$\sqrt{\text{Hz}}$ \cite{borowka2024continuous}. The sensitivity limit of RAQRs may even reach pV/cm/$\sqrt{\text{Hz}}$ obeying the standard quantum limit (SQL). Secondly, RAQRs are capable of receiving RF frequencies spanning from near direct-current to Terahertz (THz) frequencies using a single vapor cell, exhibiting broadband tunability. One or more of the RF signals to be processed may be at different but specific discrete frequencies \cite{holloway2014broadband, zhou2022theoretical, holloway2021multiple}, or at any frequency of a specific continuous range \cite{simons2021continuous, liu2022continuous, berweger2023rydberg}. Thirdly, RAQRs can directly down-convert RF signals to baseband without using any complex processing, hence significantly simplifying the receiver's structure compared to that of the conventional RF receiver. More particularly, the size of RAQRs is independent of the wavelength of RF signals. Furthermore, RAQRs are capable of acquiring precise measurements directly linked to SI without any calibration. Finally, RAQRs are omnidirectional, capable of receiving RF signals from all angular directions.

	The above-mentioned prominent features and powerful capabilities of RAQRs make them appealing in assisting classical wireless communications and sensing. However, the current studies are essentially limited to an experimental perspective from the physics community. They focused on improving the sensitivity \cite{jing2020atomic, prajapati2021enhancement, borowka2024continuous,Wu2025EnhancingCavity,Wu2025EnhancingArray}, and on developing multi-band \cite{holloway2014broadband, zhou2022theoretical, holloway2021multiple} or continuous-band detection \cite{simons2021continuous, liu2022continuous, berweger2023rydberg}. They also realize various functionalities, such as detection of amplitude \cite{sedlacek2012microwave, fan2015atom, Fancher2021Rydberg}, phase \cite{simons2019rydberg, jing2020atomic, prajapati2021enhancement}, polarization \cite{sedlacek2013atom, anderson2018vapor, anderson2021self}, modulation \cite{anderson2020atomic, holloway2019detecting, nowosielski2024warm}, spatial direction \cite{robinson2021determining}, and spatial displacement \cite{zhang2023quantum}. Furthermore, several initial applications verifying the benefits of RAQRs in assisting communications were experimentally presented in \cite{meyer2018digital, cai2023high, yuan2023rydberg, zhang2024image}. The above experimental studies have demonstrated the feasibility of RAQRs, but they fail to support theoretical studies due to the lack of a complete end-to-end signal model. A pair of emerging studies \cite{cui2024towards, Yuan2024Electromagnetic} from the communication society, introduce RAQRs to MIMO communications from signal processing and electromagnetic (EM) modeling perspectives. The former mainly emphasize their algorithmic designs based on a system model abstracted from a two-level quantum system. However, this model may be simple and inaccurate in describing a realistic RAQR. 
	Moreover, a comprehensive overview of RAQRs designed for classical wireless communications and sensing was presented in \cite{gong2024RAQRs}. In summary, the study of RAQRs for classical wireless communications and sensing is in its infancy. Clearly, a practical signal model, bridging physics and communications/sensing, is urgently needed for system design and signal processing.

	To fill this knowledge gap and unlock the full potential of RAQRs in classical wireless communications and sensing, we study the superheterodyne RAQR as a benefit of its high sensitivity and capability in both amplitude and phase detection. Explicitly, we offer the following contributions. 
	\begin{itemize}[leftmargin=2.5mm]
		\item 
		We study the typical four-level quantum scheme in depicting the realistic RAQRs, and derive a closed-form expression of the density coefficient related to the so-called $\ket{1}$\textrightarrow$\ket{2}$ transition under realistic assumptions. We associate the probe beam with this density coefficient for characterizing its amplitude and phase. The expression derived offers a precise evaluation on the transformation from the input RF signal to the output probe beam (transfer function of RAQRs). It also paves the way for jointly optimizing diverse parameters of RAQRs, e.g., the detuning and power of laser beams. 
		
		\item 
		We construct an end-to-end RAQR-aided wireless reception scheme by detailing each of the functional blocks and derive the corresponding input-output equivalent baseband signal model of the overall system. The scheme follows a realistic signal reception flow of RAQRs, offering an explicit design guidance for RAQR-aided wireless systems. The proposed signal model reveals that the RAQR imposes a gain and phase shift on the baseband transmit signal, where these factors are determined by both the atomic response and the specific photodetection scheme. The form of our signal model is consistent with that of classical RF receivers, paving the way for performing various signal processing tasks. 
		
		\item 
		We further investigate the noise sources that contaminate an RAQR, where both extrinsic and intrinsic noises are considered. The former includes the black-body radiation (BBR) induced thermal noise \cite{Fancher2021Rydberg,santamaria2022comparison}, while the latter encompasses the quantum projection noise (QPN) \cite{kitching2011atomic,fan2015atom,cox2018quantum} due to the probabilistic collapse of superposition states, the photon shot noise (PSN) \cite{Kasap2013Optoelectronics} due to the photodetection, and the intrinsic thermal noise (ITN) \cite{orfanidis2002electromagnetic} due to electronic components. Particularly, we quantifies the power of these noises at the receiver baseband, where the PSN may dominate among all noise sources depending on the photodetection scheme employed, and the QPN offers an ultimate fundamental limit of RAQRs. 
		
		\item 
		We next study the performance of RAQR-aided wireless systems, where we present the received SNR of the system for two different photodetection schemes, namely the direct incoherent optical detection and the balanced coherent optical detection. We theoretically demonstrate that the latter scheme outperforms the former scheme in terms of the received SNR. We also provide deeper insights for the received SNR in both the SQL and photon shot limit (PSL) regimes, showcasing the ultimate fundamental limit and the practical photodetection limit of RAQRs, respectively. We note that the balanced coherent optical detection scheme is capable of approaching the PSL. 
		Furthermore, we compare RAQRs to classical RF receivers, where we derive concise and useful expressions of the SNR ratio over the classical counterparts. Particularly, Doppler broadening-free RAQRs can outperform classical RF receivers in terms of the received SNR, offering an extra SNR gain of $27$ dB and $40$ dB in both the PSL and SQL regimes, respectively. 
		
		\item 
		We finally perform diverse numerical simulations to validate the proposed signal models, optimize the system parameters, characterize the performance of RAQR-aided wireless systems, and to investigate the influence of several practical impairments. Our results verify the effectiveness of our models and parameter optimization, demonstrate the superiority of RAQRs in assisting classical wireless communications and sensing, as well as facilitate the understanding of performance mismatches between theory and practice. 
	\end{itemize}

	\textit{Organization and Notations}: 
	The article is organized as follows. In Section \ref{Section:AtomicReceiverModel}, we construct the quantum response model of the RAQRs. In Section \ref{SubSection:BasebandScheme&Model}, we propose an RAQR-aided wireless reception scheme and detail the corresponding equivalent baseband signal model. In Section \ref{sec:PerformanceAnalysis}, we analyse the performance of RAQRs. We then proceed by presenting our simulation results in Section \ref{sec:Simulations}, promote helpful discussions in \ref{sec:Discussions}, and finally conclude the article in Section \ref{sec:Conclusions}. 
	The notations $\frac{\mathrm{d} \bm{\rho}}{\mathrm{d} t}$ represent the differential of $\bm{\rho}$ with respect to time; $\comm{\bm{H}}{\bm{\rho}} = \bm{H}\bm{\rho} - \bm{\rho}\bm{H}$ represents the commutator; $\left\{ \bm{\varGamma}, \bm{\rho} \right\} = \bm{\varGamma}\bm{\rho} + \bm{\rho}\bm{\varGamma}$ stands for the anticommutator; $\mathscr{R} \{ \cdot \}$ and $\mathscr{I} \{ \cdot \}$ take the real and imaginary parts of a complex number; $\text{diag}\{ \cdot \}$ denotes the diagonal of a matrix; $\chi'$ represents the derivative of $\chi$; $\hbar$ is the reduced Planck constant and $\jmath^2 = 1$; $Z_0$ and $c$ are the impedance and the speed of light in free space, respectively, $\epsilon_0$ is the vacuum permittivity, and $Z_0 = 1/(c \epsilon_0)$; $\eta_0$ and $\eta_1$ are the antenna efficiency and the quantum efficiency of the photodetector, respectively; $q$ and $a_0$ are the elementary charge and Bohr radius, respectively.

	\vspace{-0.8em}
	\section{The Quantum Response Model of RAQRs}
	\label{Section:AtomicReceiverModel}
	
	The architecture of RAQR is portrayed in Fig \ref{fig:RAQRScheme}(a). Briefly, it consists of a vapor cell containing alkali atoms, namely Cesium (Cs) or Rubidium (Rb) atoms, and a pair of laser beams, termed as probe and coupling. These laser beams counter-propagate through the vapor cell to form a spatially overlapped receive area, in which Rydberg atoms are prepared. The RAQR serves as an RF to optical converter, which is facilitated by utilizing the EIT phenomenon \cite{schlossberger2024rydberg, zhang2024rydberg, Fancher2021Rydberg, gong2024RAQRs}. When an RF signal impinging to the Rydberg atoms, a corresponding spectroscopic signal generates as a consequence of the atom-field interactions, which is further detected by a photodetector. 
	
	When only the desired RF signal is received by Rydberg atoms, the structure is termed as the standard scheme \cite{gong2024RAQRs}. Differently, when a superposition of the desired RF signal and a local oscillator (LO) alters the atomic response, the structure is superheterodyne. 
	We emphasize the superheterodyne structure of RAQRs \cite{jing2020atomic} in our receiver, as a benefit of its high sensitivity and powerful capability of receiving various modulated RF signals. We also note that our models of \eqref{eq:rho_21}, \eqref{eq:OutputProbeSig}, \eqref{eq:OutputProbeSig_Baseband} in Section \ref{SubSection:AtomicTransitionModel} and Section \ref{SubSection:TransferFunction} are also applicable to the standard structure (without the LO) that exploits the Autler–Townes splitting (ATS) for RF signal detection.

	\vspace{-1em}
	\subsection{The Electron Transition Model of RAQRs}
	\label{SubSection:AtomicTransitionModel}
	
\begin{figure}[t!]
		\centering
		\includegraphics[width=0.49\textwidth]{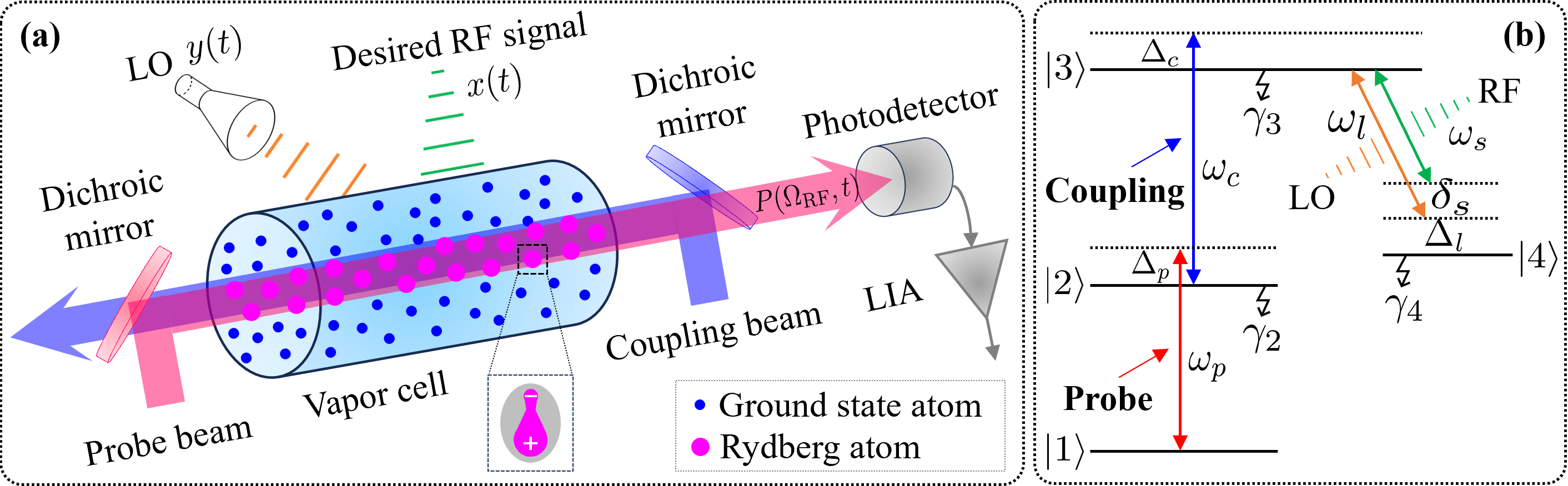}
		\caption{(a) The superheterodyne structure of RAQRs, and (b) its corresponding four-level scheme.}
		\vspace{-1.5em}
		\label{fig:RAQRScheme}
	\end{figure}

	The electron transitions happening to Rydberg atoms in the receive area of the vapor cell may be described by a four-level ladder scheme, as shown in Fig. \ref{fig:RAQRScheme}(b). It can be viewed in two processes: Rydberg atom preparation and RF signal coupling.

	\textit{Rydberg atom preparation}: 
	The electron transition between the ground state $\ket{1}$ and the excited state $\ket{2}$ is (near-) resonantly coupled with the probe beam having its Rabi frequency\footnote{The Rabi frequency describes the frequency of an electron fluctuating between two energy levels when imposing an oscillating EM field. It is directly related to the amplitude of this EM field \cite{Fancher2021Rydberg}.}, angular frequency, and detuning frequency\footnote{The detuning frequency $\Delta_{p,c}$ represents the frequency difference between the resonant angular frequency $\omega_{i-1,i}, i=2, 3$ of two energy levels and the actual angular frequency $\omega_{p,c}$ of the EM field coupled to these energy levels.} of $\{ \Omega_p, \omega_{p}, \Delta_{p} \}$. This excited atom is further driven to a specific Rydberg state $\ket{3}$ by the coupling beam having its Rabi frequency, angular frequency, and detuning frequency of $\{ \Omega_c, \omega_{c}, \Delta_{c} \}$. The probe/coupling beams are resonant when $\Delta_{p,c} = 0$ and are near resonant when $\Delta_{p,c}$ spanning a small frequency range within the response range of the energy level.

	\textit{RF signal coupling}: For the standard structure \cite{gong2024RAQRs} (without the LO) of RAQRs, the desired RF signal of $\{ \Omega_x, \omega_{x}, \Delta_{x} \}$ directly drives the so-called Rydberg states $\ket{3}$ and $\ket{4}$. 
	By contrast, for the superheterodyne structure of RAQRs, the LO of $\{ \Omega_l, \omega_{l}, \Delta_{l} \}$ (near-) resonantly excites the transition between $\ket{3}$ and $\ket{4}$. 
	The impinging RF signal to be detected produces a coupling $\Omega_x \cos{(2 \pi f_{\delta} t + \theta_{\delta})}$, which has a frequency offset of $f_{\delta}$ and a phase offset of $\theta_{\delta}$ relative to the LO.  Furthermore, the RAQR requires $\Omega_{x} \ll \Omega_{l}$, implying that the impinging RF is weak compared to the LO \cite{jing2020atomic}.

	To elaborate a little further, typically, the Lindblad master equation is applied for characterizing the dynamics of the four-level transition scheme, which is given by \cite{auzinsh2010optically} 
	\begin{align}
		\label{eq:Lindbladian}
		\frac{\mathrm{d} \bm{\rho}}{\mathrm{d} t} 
		= - \jmath \comm{\bm{H}}{\bm{\rho}} - \frac{1}{2} \left\{ \bm{\varGamma}, \bm{\rho} \right\} + \bm{\varLambda},
	\end{align}
	where $\bm{H}$, $\bm{\varGamma}$ and $\bm{\varLambda}$ are the Hamiltonian, the relaxation matrix, and the decay matrix, respectively. They are given by \cite{auzinsh2010optically}
	\begin{align}
		\label{eq:Hamiltonian}
		\bm{H} &= \left[ {\begin{array}{*{20}{c}}
				0 & {\frac{{{\Omega _p}}}{2}} & 0 & 0\\
				{\frac{{{\Omega _p}}}{2}} & {{\Delta _p}} & {\frac{{{\Omega _c}}}{2}} & 0\\
				0 & {\frac{{{\Omega _c}}}{2}} & {{\Delta _p} + {\Delta _c}} & {\frac{{{\Omega _{\text{RF}}}}}{2}}\\
				0 & 0 & {\frac{{{\Omega _{\text{RF}}}}}{2}} & {{\Delta _p} + {\Delta _c} + {\Delta _l}}
		\end{array}} \right],
	\end{align}
	$\bm{\varGamma} = \text{diag} \{ \gamma, \gamma + \gamma_2, \gamma + \gamma_3 + \gamma_c, \gamma + \gamma_4 \}$, and $\bm{\varLambda} = \text{diag} \{ \gamma + \gamma_2 \rho_{22} + \gamma_4 \rho_{44}, \gamma_3 \rho_{33}, 0, 0 \}$, where $\gamma_{i}$ is the spontaneous decay rate of the $i$-th level. In \eqref{eq:Hamiltonian}, $\Omega_{\text{RF}} \in \{ \Omega_x, \Omega_l + \Omega_x \cos{(2 \pi f_{\delta} t + \theta_{\delta})}\footnote{The linear form of this equation is discussed in Section \ref{subsec:RFsignals}.} \}$ depending on the structure selected. 
	To elaborate a little further, $\gamma$ and $\gamma_c$ are the relaxation rates related to the atomic transition effect and the atomic collision effect, respectively. For simplicity, we assume $\gamma = \gamma_c = 0$. We also assume that the decay rates of $\ket{3}$ and $\ket{4}$ are comparatively low so that they can be reasonably ignored \cite{jing2020atomic}.

	Based on the above assumptions, we can formulate a closed-form expression of the steady-state (when $\frac{\mathrm{d} \bm{\rho}}{\mathrm{d} t} = 0$) solution of the density matrix $\bm{\rho}$. More particularly, we are interested in $\rho_{21}$ of $\bm{\rho}$, since it is associated with the probe beam to be measured. Specifically, we derive $\rho_{21}$ as follows 
	\begin{align}
		\nonumber
		&\rho_{21} (\Omega_{\text{RF}}) = {\Omega _p} \times \\
		\label{eq:rho_21} 
		&\frac{ {A_1}\Omega _{{\rm{RF}}}^4 + {A_2}\Omega _{{\rm{RF}}}^2 + {A_3} - \jmath \left( {B_1}\Omega _{{\rm{RF}}}^4 + {B_2}\Omega _{{\rm{RF}}}^2 + {B_3} \right) }{ {C_1}\Omega _{{\rm{RF}}}^4 + {C_2}\Omega _{{\rm{RF}}}^2 + {C_3} },
	\end{align}
	where the coefficients $A_1, A_2, A_3$, $B_1, B_2, B_3$, $C_1, C_2, C_3$ are presented in Appendix \ref{DensityMatrixCoeffs}. We note that these coefficients can be simplified in the following special cases: 
	\begin{enumerate}[label={\em {C\arabic*}}, leftmargin=1.7em, labelindent=0pt, itemindent = 0em]
		\item \label{itm:C1} 
		The probe beam is near resonant, while the coupling beam and LO are perfectly resonant, i.e., $\Delta_{p} \neq 0$, $\Delta_{c/l} = 0$.
		\item \label{itm:C2} 
		The coupling beam is near resonant, while the probe beam and LO are perfectly resonant, i.e., $\Delta_{c} \neq 0$, $\Delta_{p/l} = 0$.
		\item \label{itm:C3} 
		The LO is near resonant, while the probe and coupling beams are perfectly resonant, i.e., $\Delta_{l} \neq 0$, $\Delta_{p/c} = 0$.
	\end{enumerate}
	The corresponding coefficient expressions can be formulated by substituting zero into the corresponding location of $\Delta_{p}$, $\Delta_{c}$ and $\Delta_{l}$, respectively, which are omitted here.

	\textit{\textbf{Remark 1:}
	\ding{172}
	The above three cases may become important when independently optimizing the three detuning frequencies to obtain an enhanced sensitivity of the RAQRs, as studied in \cite{wu2023theoretical, wu2024atomic}. It is worth noting that these studies are no longer relevant however, when the three detuning frequencies have to be jointly optimized. However, their joint optimization will potentially offer an improved sensitivity over those of \cite{wu2023theoretical, wu2024atomic} (as seen in Fig. \ref{fig:ParameterOptimization}(f) in Section \ref{subsec:OP}). Our expression \eqref{eq:rho_21} is more general than that of previous studies and can be employed for jointly optimizing detuning frequencies. A simple optimization of the RAQR parameters by exploiting \eqref{eq:rho_21} can be seen in Section \ref{subsec:OP}.}

	\textit{\ding{173} 
	We also note that our model \eqref{eq:rho_21} does not rely on the so-called weak probe beam assumption ($\Omega_p \ll \Omega_c$), as widely assumed in the physics community \cite{borowka2022sensitivity}.}

	\vspace{-1em}
	\subsection{The RF-to-Optical Transformation Model of RAQR}
	\label{SubSection:TransferFunction}
	
	The RAQR realizes an RF-to-optical transformation that depends on the electron transition model presented in Section \ref{SubSection:AtomicTransitionModel}. The desired RF signal serves as the input of the RAQR and the probe beam acts as its output. To begin with, we assume that the laser beams are Gaussian beams, and express the probe beam at the access area of the atomic vapor cell as 
	\begin{align}
		\label{eq:InputProbeSig}
		P_{0}(t) 
		&= \sqrt{ 2 \mathcal{P}_{0} } \cos( 2 \pi f_{p} t + \phi_{0} ) = \sqrt{2} \mathscr{R} \left\{ P_{0b} e^{\jmath 2 \pi f_p t} \right\},
	\end{align}
	where $\mathcal{P}_{0}$, $f_{p}$, and $\phi_{0}$ are the power, frequency, and phase of the input probe beam, respectively. The power is given by $\mathcal{P}_{0} = \frac{\pi c \epsilon_0}{8 \ln {2}}  F_p^2 \left| U_{0} \right|^{2}$ \cite{robinson2021atomic}, where $F_{p}$ represents the full width at half maximum (FWHM) of the probe beam and $U_{0}$ is the amplitude of this electric field. Furthermore, $P_{0b} = \sqrt{\mathcal{P}_{0}} e^{\jmath \phi_0}$ denotes the equivalent baseband signal of the passband probe beam. After propagating through the vapor cell, the probe beam is influenced by Rydberg atoms both in terms of its amplitude and phase at the output of the vapor cell. Let us denote the amplitude and phase of the output probe beam by $U_{p} (\Omega_{\text{RF}})$ and $\phi_{p} (\Omega_{\text{RF}})$, respectively. They can be associated with their input counterparts formulated as \cite{auzinsh2010optically}, \cite{meyer2021optimal} 
	\begin{align}
		\label{eq:AmplitudeRelation}
		U_{p} (\Omega_{\text{RF}}) 
		&= U_{0} e^{ - \frac{\pi d}{\lambda_p} \mathscr{I} \{\chi (\Omega_{\text{RF}}) \}  }, \\
		\label{eq:PhaseRelation}
		\phi_{p} (\Omega_{\text{RF}})  
		&= \phi_{0} + \tfrac{\pi d}{\lambda_p} \mathscr{R} \{\chi (\Omega_{\text{RF}}) \}, 
	\end{align}
	where $\lambda_p$ is the wavelength of the probe laser, $d$ denotes the length of the vapor cell, and $\chi (\Omega_{\text{RF}})$ is the susceptibility of the atomic vapor medium. We note that \eqref{eq:AmplitudeRelation} is known as the Beer–Lambert law.

	In \eqref{eq:AmplitudeRelation} and \eqref{eq:PhaseRelation}, the macroscopic susceptibility is associated with the microscopic coherence in \eqref{eq:rho_21} through 
	\begin{align}
		\label{eq:susceptibility_density_relation}
		\chi(\Omega_{\text{RF}}) = - \tfrac{2 N_0 \mu_{12}^{2}}{\epsilon_0 \hbar \Omega_p} \rho_{21}(\Omega_{\text{RF}}), 
	\end{align}
	where $N_0$ is the atomic density in the vapor cell and $\mu_{12}$ is the dipole moment of transition of $\ket{1} \rightarrow \ket{2}$. Therefore, the output probe beam (passband) can be formulated as 
	\begin{align}
		\nonumber
		P(\Omega_{\text{RF}}, t) 
		&= \sqrt{ 2\mathcal{P}_{1} (\Omega_{\text{RF}}) } \cos \left( 2 \pi f_{p} t + \phi_{p} (\Omega_{\text{RF}}) \right) \\
		\label{eq:OutputProbeSig}
		&= \sqrt{2} \mathscr{R} \left\{ P_b(\Omega_{\text{RF}}, t) e^{\jmath 2 \pi f_p t} \right\},
	\end{align}
	where $\mathcal{P}_{1} (\Omega_{\text{RF}}) = \frac{\pi c \epsilon_0}{8 \ln {2}}  F_p^2 \left| U_{p}(\Omega_{\text{RF}}) \right|^{2}$ represents the power of the output probe beam and $P_b(\Omega_{\text{RF}}, t) = \sqrt{\mathcal{P}_{1} (\Omega_{\text{RF}})} e^{\jmath \phi_p (\Omega_{\text{RF}})}$ is the equivalent baseband signal, specifying 
	\begin{align}
		\label{eq:OutputProbeSig_Baseband}
		P_b(\Omega_{\text{RF}}, t) 
		= \sqrt{\mathcal{P}_{1} (\Omega_{\text{RF}})}
		e^{\jmath \phi_{0} } e^{\jmath \frac{\pi d}{\lambda_p} \mathscr{R} \{\chi (\Omega_{\text{RF}}) \} }. 
	\end{align}
	\textit{\textbf{Remark 2:}
	We note that \eqref{eq:OutputProbeSig} and \eqref{eq:OutputProbeSig_Baseband} are nonlinear functions due to the nonlinear susceptibility. Additionally, we emphasize that \eqref{eq:OutputProbeSig} and \eqref{eq:OutputProbeSig_Baseband} are valid for both the standard and superheterodyne structures \cite{gong2024RAQRs} in numerical simulations. However, in practice, the ATS spectroscopic signal of the standard structure is only obtained by scanning the frequency of the coupling (or probe) beam in a specific range. This leads to a time-varying detuning $\Delta_c$ (or $\Delta_p$), which increases the difficulty to obtain an explicit representation of \eqref{eq:OutputProbeSig} and \eqref{eq:OutputProbeSig_Baseband} for the standard scheme. By contrast, the superheterodyne structure can be configured with fixed $\Delta_{p,c,l}$ without requiring any probe/coupling scanning.}

	\vspace{-1em}
	\section{The Proposed RAQR-Aided Wireless Scheme and its Signal Model}
	\label{SubSection:BasebandScheme&Model}
	
\begin{figure*}[t!]
		\centering
		\includegraphics[width=0.9\textwidth]{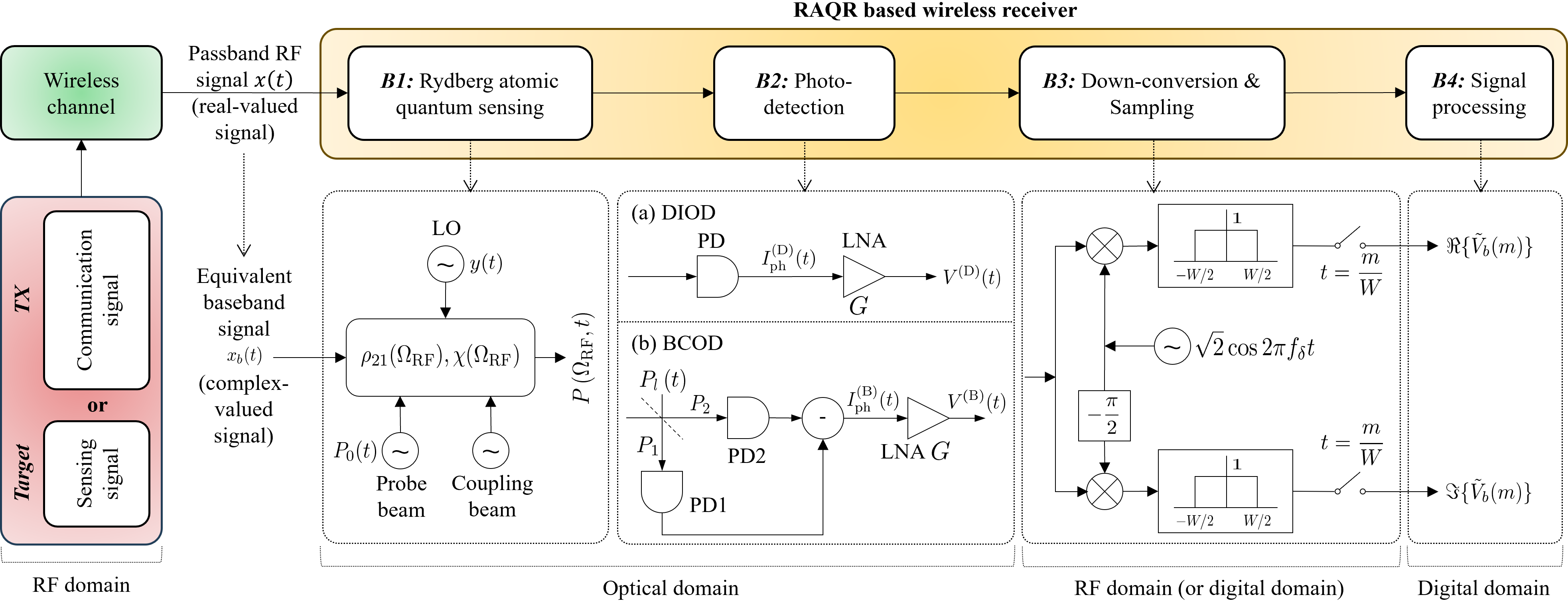}
		\caption{Illustration of the RAQR-aided wireless receiver.}
		\vspace{-1.5em}
		\label{fig:ReceiveScheme}
	\end{figure*}

	Upon incorporating the RAQR into classical wireless communications and sensing to form an RAQR-aided wireless system, we investigate the end-to-end system from an equivalent baseband signal perspective. In the following contents, we first present a detailed block diagram to illustrate the end-to-end reception process of the RAQR-aided wireless system, and then derive the corresponding end-to-end equivalent baseband signal model. The block diagram of the RAQR-aided wireless system is presented in Fig. \ref{fig:ReceiveScheme}.

	\vspace{-1em}
	\subsection{RF Signal to be Detected}
	\label{subsec:RFsignals}
	
	The RF signals to be detected may be constituted by a modulated communication signal or a sensing signal (such as radar, sonar, and ultrasound signals). We assume that these signals propagate in form of plane waves. Then we formulate the unified mathematical model of the received RF signal as  
	\begin{align}
		\label{eq:RFsignal}
		\hspace{-0.5em} x (t) 
		= \sqrt{2 \mathcal{P}_{x}} \cos ( 2 \pi f_c t + \theta_x ) 
		= \sqrt{2} \mathscr{R} \left\{ x_b (t) e^{\jmath 2 \pi f_c t} \right\}, 
	\end{align}
	where $\mathcal{P}_{x}$ and $\theta_x$ are the power and phase of the RF signal. The power is given by $\mathcal{P}_{x} = \frac{1}{2} c \epsilon_0 A_e \left| U_{x} \right|^{2}$, where $A_e$ is the effective receiver aperture of the RAQR and $U_x$ is the amplitude of the RF signal. Furthermore, $x_b (t) = \sqrt{\mathcal{P}_{x}} e^{\jmath \theta_x}$ represents the equivalent baseband signal. The specific forms of practical communication and sensing signals connecting the transmitter and the RAQR are detailed in Section \ref{subsec:SignalModel}. 
	
	Let us further denote the LO impinging to the RAQR as 
	\begin{align}
		\label{eq:LO}
		y (t) = \sqrt{2\mathcal{P}_{y}} \cos ( 2 \pi f_l t + \theta_{y} ) = \sqrt{2} \mathscr{R} \left\{ y_b (t) e^{\jmath 2 \pi f_l t} \right\}, 
	\end{align} 
	where $\mathcal{P}_{y}$ and $\theta_{y}$ represent the power and phase, and $y_{b} (t) = \sqrt{\mathcal{P}_{y}} e^{\jmath \theta_y}$ is the equivalent baseband signal.  The power is given by $\mathcal{P}_{y} = \frac{1}{2} c \epsilon_0 A_{e} \left| U_{y} \right|^{2}$, where $U_{y}$ is the amplitude of the LO. 
	
	Recall that $f_{\delta} = f_c - f_l$ and $\theta_{\delta} = \theta_{x} - \theta_y$ are the frequency difference and phase difference between the desired RF signal and the LO. 
	Then the superposition of the desired RF signal and the LO is received by Rydberg atoms and it is given by 
	\begin{align}
		\nonumber
		z (t) = x (t) + y (t) 
		&= \sqrt{2} \mathscr{R} \left\{ \left[ x_b (t) e^{\jmath 2 \pi f_{\delta} t} + y_b (t) \right] e^{\jmath 2 \pi f_l t} \right\} \\
		\label{eq:PassbandRFplusLO}
		&\overset{(a)}{\approx} \sqrt{2\mathcal{P}_{z}} \cos( 2 \pi f_l t + \theta_y ),
	\end{align}
	where $\mathcal{P}_{z} = \frac{1}{2} c \epsilon_0 A_{e} \left| U_{z} \right|^{2}$ represents the power of $z(t)$. In the expression of $\mathcal{P}_{z}$, we have the amplitude $U_z$ formulated as
	\begin{align}
		\nonumber
		U_z &= \sqrt{ \left| U_{x} \right|^{2} + \left| U_{y} \right|^{2} + 2 U_x U_y \cos( 2 \pi f_{\delta} t + \theta_{\delta} ) } \\
		\label{eq:Amp_PassbandRFplusLO}
		&\overset{(b)}{\approx} U_y + U_x \cos( 2 \pi f_{\delta} t + \theta_{\delta} ).
	\end{align} 
	See Appendix \ref{RFsuperpositionProof} for the proofs of \eqref{eq:PassbandRFplusLO} and \eqref{eq:Amp_PassbandRFplusLO}.
	
	Furthermore, we obey the linear relationship between the Rabi frequency and amplitude of the RF signal in the form of $\Omega_{\text{RF}} = \frac{ \mu_{34} }{ \hbar } U_z$, $\Omega_y = \frac{ \mu_{34} }{ \hbar } U_y$, and $\Omega_x = \frac{ \mu_{34} }{ \hbar } U_x$, where $\mu_{34}$ represents the dipole moment corresponding to the electron transition between Rydberg states. The separation form of \eqref{eq:Amp_PassbandRFplusLO} enables the expression of the superimposed Rabi frequency $\Omega_{\text{RF}} = \Omega_l + \Omega_x \cos{(2 \pi f_{\delta} t + \theta_{\delta} )}$ of \eqref{eq:Hamiltonian} for superheterodyne structures presented in Section \ref{SubSection:AtomicTransitionModel}.

	\vspace{-1em}
	\subsection{Modeling of RAQR Based Wireless Receivers}
	\label{subsec:RAQR}
	
	For the RAQR based wireless receiver, we characterize its four blocks as follows.

	\subsubsection{Rydberg Atomic Quantum Sensing and Photodetection}
	For the receiver block of Rydberg atomic quantum sensing, the electron transition model has been constructed in \eqref{eq:rho_21} of Section \ref{SubSection:AtomicTransitionModel} and the RF-to-optical transformation model has been provided in \eqref{eq:OutputProbeSig} of Section \ref{SubSection:TransferFunction}. In both equations, the $\Omega_{\text{RF}}$ is a linear superimposition of both the LO and desired RF signal related components, namely $\Omega_{\text{RF}} = \Omega_l + \Omega_x \cos{(2 \pi f_{\delta} t + \theta_{\delta} )}$, as proved in Section \ref{subsec:RFsignals}. 
	
	Based on these results, we study two typical photodetection schemes, namely the direct incoherent optical detection (DIOD) and the balanced coherent optical detection (BCOD) \cite{haus2012electromagnetic}, and present their corresponding nonlinear outputs and associated linear approximations.

	\textbf{DIOD}: 
	In this scheme, the output probe beam is directly received, integrated over a period, and then generates a photocurrent according to the incident beam power, as modeled in the first scheme of ``\textsl{\textbf{B2}}: Photodetection" of Fig. \ref{fig:ReceiveScheme}. We first derive the power of the equivalent baseband signal relying on \eqref{eq:OutputProbeSig}, which is further transformed to an output photocurrent according to the relationship $I_{\text{ph}}^{\text{(D)}} (t) = \alpha |P_{b} (\Omega_{\text{RF}})|^{2}$, where ${\alpha \triangleq \frac{{\eta_1 q}}{{2 \pi \hbar f_p }}}$ denotes the photodetector responsivity. A voltage is produced by the photocurrent across a load (we neglect it by assuming a unit value), which is further amplified by a low noise amplifier (LNA) (gain $G$). Consequently, the amplified voltage is expressed as 
	\begin{align}
		\nonumber
		&V^{\text{(D)}}(t) 
		= \sqrt{G} I_{\text{ph}}^{\text{(D)}} (t) 
		= \sqrt{G} \alpha \mathcal{P}_{1} (\Omega_{\rm{RF}}) 
		\overset{(c)}{\approx} \sqrt{G} \alpha \mathcal{P}_{1} (\Omega_{l}) \\
		\label{eq:PhotodetectorOutputDIOD}
		&\qquad \times \big[ 1 - 2 \kappa_1 (\Omega_l)  \cos {\varphi_1({\Omega _l})} U_x \cos \left( 2\pi f_{\delta} t + \theta_{\delta} \right) \big], 
	\end{align}
	where we have 
	\begin{align}
		\label{eq:kappa_1}
		&\kappa_1 (\Omega_l) = \tfrac{{\pi d \mu_{34}}}{ {\lambda _p} \hbar } \left| {\mathscr I}\{ \chi '({\Omega _l})\} \right|, \\
		\label{eq:varphi_1}
		&\varphi_1({\Omega _l}) = \arccos{ \frac{ {\mathscr I}\{ \chi '({\Omega _l})\} }{ \left| {\mathscr I}\{ \chi '({\Omega _l})\} \right| } }.
	\end{align}
	The approximation in \eqref{eq:PhotodetectorOutputDIOD} is derived using Taylor series expansion in the vicinity of a given value of $\Omega_{\text{RF}} = \Omega_l$.

	\textbf{BCOD}: 
	A strong local optical beam exists in this scheme, as portrayed in the second scheme of ``\textsl{\textbf{B2}}: Photodetection" of Fig. \ref{fig:ReceiveScheme}. Exploiting such a local optical source can help to suppress the thermal noise generated by the remaining electronic components.  
	We assume that its carrier frequency and FWHM are identical to those of the probe beam. Then we can express the local optical beam in the form of $P_{l} (t) = \sqrt{2\mathcal{P}_{l}} \cos ( 2 \pi f_p t + \phi_l ) = \sqrt{2} \mathscr{R} \left\{ P_{lb} (t) e^{\jmath 2 \pi f_p t} \right\}$, where $\mathcal{P}_{l} = \frac{\pi c \epsilon_0}{8 \ln {2}}  F_p^2 \left| U_{l} \right|^{2}$ and $P_{lb} (t) = \sqrt{\mathcal{P}_{l}} e^{\jmath \phi_l}$.  Furthermore, the probe beam and the local optical beam are combined to form two distinct optical beams, namely ${P_1} = \frac{1}{{\sqrt 2 }}\left[ {{P_l}\left( t \right) - P\left( {{\Omega _{{\rm{RF}}}},t} \right)} \right]$ and ${P_2} = \frac{1}{{\sqrt 2 }}\left[ {{P_l}\left( t \right) + P\left( {{\Omega _{{\rm{RF}}}},t} \right)} \right]$. They are detected by two photodetectors, respectively. The photocurrents generated are subtracted to obtain an output photocurrent, given by $I_{\text{ph}}^{\text{(B)}}(t) = \alpha \left[ {{P_{lb}}\left( t \right){P_b^*}\left( {{\Omega _{{\rm{RF}}}},t} \right) + P_{lb}^*\left( t \right) P_b \left( {{\Omega _{{\rm{RF}}}},t} \right)} \right]$ \cite{haus2012electromagnetic}. Upon using the same resistance and LNA as in DIOD, we formulate the output voltage of the BCOD scheme as follows
	\begin{align}
		\nonumber
		\hspace{-1em} V^{\text{(B)}}(t) &= \sqrt{G} I_{\text{ph}}^{\text{(B)}} (t) 
		= 2 \sqrt{G} \alpha \sqrt{\mathcal{P}_{l} \mathcal{P}_{1} (\Omega_{\rm{RF}})}
		\cos \left( {\phi _l} - {\phi _p}({\Omega _{{\rm{RF}}}}) \right) \\
		\nonumber
		&\overset{(d)}{\approx} 2 \sqrt{G} \alpha \sqrt{\mathcal{P}_{l} \mathcal{P}_{1} (\Omega_{l})} 
		\Big[ \cos \left( {\phi _l} - {\phi _p}({\Omega _l}) \right) \Big. \\
		\label{eq:PhotodetectorOutputBCOD}
		&\Big. \;\;\;\;\;\; - \kappa_2 (\Omega_l) \cos {\varphi_2({\Omega _l})} U_x \cos \left( {2\pi {f_\delta }t + {\theta _\delta }} \right) \Big], 
	\end{align}
	where we have 
	\begin{align}
		\label{eq:kappa2}
		&\kappa_2 (\Omega_l) = \tfrac{\pi d \mu_{34}}{\lambda_p \hbar } \sqrt {{{\left[ {{\mathscr I}\{ \chi '({\Omega _l})\} } \right]}^2} + {{\left[ {{\mathscr R}\{ \chi '({\Omega _l})\} } \right]}^2}}, \\
		\label{eq:varphi}
		&{\varphi_2({\Omega _l})} = {\phi _l} - {\phi _p}({\Omega _l}) + {\psi _p}({\Omega _l}), \\
		\label{eq:psi_p}
		&\psi_p(\Omega_{l}) = \arccos \frac{ {{\mathscr I}\{ \chi '({\Omega _l})\} } }{ \sqrt {{{\left[ {{\mathscr I}\{ \chi '({\Omega _l})\} } \right]}^2} + {{\left[ {{\mathscr R}\{ \chi '({\Omega _l})\} } \right]}^2}} }, \\
		\nonumber
		&\mathscr{R} \{ \chi' \left( \Omega_{l} \right) \} 
		= - \tfrac{4 N_0 \mu_{12}^{2}}{\epsilon_0 \hbar} {\Omega _l} \left[ \frac{ {2{A_1}\Omega _l^2 + {A_2}} }{{{C_1}\Omega _{{l}}^4 + {C_2}\Omega _{{l}}^2 + {C_3}}} \right. \\
		\label{eq:Chi_1_Real}
		&\left. \qquad \quad \;\;\; - \frac{ \left( {{A_1}\Omega _{{l}}^4 + {A_2}\Omega _{{l}}^2 + {A_3}} \right) \left( {2{C_1}\Omega _l^2 + {C_2}} \right)}{{{{\left( {{C_1}\Omega _{{l}}^4 + {C_2}\Omega _{{l}}^2 + {C_3}} \right)}^2}}} \right], \\
		\nonumber
		&\mathscr{I} \{ {\chi'} \left( \Omega_{l} \right) \}
		= \tfrac{4 N_0 \mu_{12}^{2}}{\epsilon_0 \hbar} {\Omega _l} \left[ \frac{ {2{B_1}\Omega _l^2 + {B_2}} }{{{C_1}\Omega _{{l}}^4 + {C_2}\Omega _{{l}}^2 + {C_3}}} \right. \\
		\label{eq:Chi_1_Imag}
		&\left. \qquad \quad \;\;\; - \frac{ \left( {{B_1}\Omega _{{l}}^4 + {B_2}\Omega _{{l}}^2 + {B_3}} \right) \left( {2{C_1}\Omega _l^2 + {C_2}} \right)}{{{{\left( {{C_1}\Omega _{{l}}^4 + {C_2}\Omega _{{l}}^2 + {C_3}} \right)}^2}}} \right]. 
	\end{align}
	See Appendix \ref{BCODOutputSignalExpressionProof} for the proof of $(d)$.

	\textit{\textbf{Remark 3:}
	\ding{172} As seen from both \eqref{eq:PhotodetectorOutputDIOD} and \eqref{eq:PhotodetectorOutputBCOD}, each output signal contains a ``direct current (DC)" independent of the RF signal and an ``alternating current (AC)" influenced by the strength, frequency, and phase of the desired RF signal. All information of the desired RF signal is embedded into the amplitude of the AC component, allowing both amplitude and phase recovery of the desired RF signal.} 
	
	\textit{\ding{173} We note that \eqref{eq:PhotodetectorOutputDIOD} and \eqref{eq:PhotodetectorOutputBCOD} are more generalized than the model of previous studies, e.g., \cite{jing2020atomic}. Our models characterize the phase \eqref{eq:PhaseRelation} of the output probe beam in addition to the amplitude \eqref{eq:AmplitudeRelation}, which is not addressed previously. This restricts previous models to the special case of $\Delta_{p,c,l} = 0$ (\eqref{eq:PhaseRelation} becomes zero). In other words, once $\Delta_{p,c,l} \ne 0$ due to practical non-idealities or specific optimizations (seen in Section \ref{subsec:OP}), employing previous models may degrade the recovery accuracy without addressing the non-zero phase \eqref{eq:PhaseRelation}.} 
	
	\textit{\ding{174} We further emphasize that the models of previous studies, e.g., \cite{jing2020atomic,wu2023theoretical, wu2024atomic}, rely on their linear approximations of either the nonlinear output probe beam $P \left( \Omega_{\text{RF}}, t \right)$ or of the nonlinear susceptibility $\chi(\Omega_{\text{RF}})$. Their linear approximations may be prematurely harnessed for a holistic wireless receiver, because the approximation error may be further spread in the remaining stages, e.g. photodetection. Instead, we apply a linear approximation to the output of the photodetector.}

	\subsubsection{Down-conversion \& Sampling}
	
	The output voltage of the photodetector is forwarded to a homodyne receiver (HR) to realize the down-conversion, as shown in ``\textsl{\textbf{B3}}: Down-conversion \& Sampling" of Fig. \ref{fig:ReceiveScheme}. After lowpass filtering, the RF-independent DC is eliminated and the output of the HR only includes the RF-dependent AC.

	Based on \eqref{eq:PhotodetectorOutputDIOD} and \eqref{eq:PhotodetectorOutputBCOD}, we can extract the time-varying component of $V^{\text{(D)}}(t)$ and $V^{\text{(B)}}(t)$ in a unified form as follows 
	\begin{align}
		\nonumber
		\tilde{V}(t) &= 
		2 \sqrt{G} \alpha \widetilde{ \mathcal{P} }
		\kappa (\Omega_l) \cos {\varphi ({\Omega _l})} U_x \cos \left( {2\pi f_{\delta} t + \theta_{\delta} } \right) \\
		\label{eq:PhotodetectorOutput_Varying}
		&= \sqrt{2} \mathscr{R} \left\{ \tilde{V}_b(t) e^{\jmath 2\pi f_{\delta} t}
		\right\},   
	\end{align}
	where we have $\widetilde{ \mathcal{P} } \in \{ \mathcal{P}_{1} (\Omega_{l}), \sqrt{\mathcal{P}_{l} \mathcal{P}_{1} (\Omega_{l})} \}$, $\kappa (\Omega_l) \in \{ \kappa_1 (\Omega_l), \kappa_2 (\Omega_l) \}$, and $\varphi ({\Omega _l}) \in \{ \varphi_1 ({\Omega _l}), \varphi_2 ({\Omega _l}) \}$ associated with the DIOD and the BCOD, respectively. By exploiting the relationship of $U_x = \sqrt{ \frac{2 {\cal P}_x}{c \epsilon_0 A_e} }$, the equivalent baseband signal in \eqref{eq:PhotodetectorOutput_Varying} is formulated as  
	\begin{align}
		\nonumber
		\tilde{V}_b(t) 
		&= \frac{2 \sqrt{G} \alpha }{ \sqrt{ c \epsilon_0 A_e } } \widetilde{ \mathcal{P} } \kappa (\Omega_l) \cos {\varphi ({\Omega _l})}
		e^{- \jmath \theta_{y}} 
		\sqrt{\mathcal{P}_{x}} e^{\jmath \theta_{x}} \\
		\label{eq:Baseband_Varying}
		&\triangleq \sqrt{ \frac{\varrho}{A_e} } \Phi x_b(t), 
	\end{align}
	where $\varrho$ and $\Phi$ denote the gain and phase of RAQR, respectively, which are expressed as 
	\begin{align}
		\label{eq:Gain}
		\varrho &= 4 Z_0 \alpha^2 G 
		\begin{cases}
			\mathcal{P}_{1}^{2} (\Omega_{l}) \kappa_1^2 (\Omega_l), & \text{DIOD},\\
			\mathcal{P}_{l} \mathcal{P}_{1} (\Omega_{l}) \kappa_2^2 (\Omega_l), & \text{BCOD},
		\end{cases} \\
		\label{eq:Phase}
		\Phi &= \begin{cases}
			\frac{ e^{ - \jmath \left( {\theta _y} - \varphi_1 ({\Omega _l}) \right) } }{2} + \frac{ e^{ - \jmath \left( {\theta _y} + \varphi_1 ({\Omega _l}) \right) } }{2}, & \text{DIOD},\\
			\frac{ e^{ - \jmath \left( {\theta _y} - \varphi_2 ({\Omega _l}) \right) } }{2} + \frac{ e^{ - \jmath \left( {\theta _y} + \varphi_2 ({\Omega _l}) \right) } }{2}, & \text{BCOD}.
		\end{cases}
	\end{align}

	Sampling both sides of \eqref{eq:Baseband_Varying} at multiples of $1/W$ as shown in ``\textsl{\textbf{B3}}: Down-conversion \& Sampling" of Fig. \ref{fig:ReceiveScheme}, we have the sampled output as
	\begin{align}
		\label{eq:SampledOutput}
		\tilde{V}_b (m) 
		&= \sqrt{ \frac{\varrho}{A_e} } \Phi x_b(m). 
	\end{align}

	\textit{\textbf{Remark 4:}
	We note that ${\kappa _1}({\Omega _l}) \le {\kappa _2}({\Omega _l})$, where the equality holds if and only if $\Delta_{p,c,l} = 0$, representing the ``no detuning" case of both laser beams and the LO. Upon exploiting $\mathcal{P}_{l} \gg \mathcal{P}_{1}({\Omega _l})$, we arrive at the result of ${\varrho _{{\rm{DIOD}}}} < {\varrho _{{\rm{BCOD}}}}$.}

	\vspace{-1em}
	\subsection{Signal Model of RAQR-Aided Wireless Systems}
	\label{subsec:SignalModel}

	\subsubsection{Equivalent Baseband Signal Model for Communications} 
	We consider a conventional RF transmitter that relies on an in-phase and quadrature-phase (IQ) modulation scheme \cite{Tse2005Fundamentals}. 
	The digital baseband signal is interpolated to an analog baseband signal via the sinc function obeying the sampling theorem. The interpolated signal is thus expressed as $s_b (t) = \sum_{n} s_b (n) \text{sinc}(Wt - n)$, where $s_b (n)$ denotes the $n$-th sample, $W$ represents the sampling rate, and $\text{sinc}(t) = \sin(\pi t) / (\pi t)$. The analog baseband signal is then upconverted to its passband counterpart $s(t) = \sqrt{2\mathcal{P}_{s}} \cos ( 2 \pi f_c t + \theta_s ) = \sqrt{2} \mathscr{R} \left\{s_b (t) e^{\jmath 2 \pi f_c t} \right\}$, where $\mathcal{P}_{s}$ is the transmit power of $s(t)$. Specifically, $\mathcal{P}_{s} = \frac{1}{2} c \epsilon_0 A_s \left| U_{s} \right|^{2}$, where $A_s$ represents the effective transmit aperture and $U_s$ is the amplitude of the RF signal. Furthermore, $s_b (t) = \sqrt{\mathcal{P}_{s}} e^{\jmath \theta_s}$ represents the equivalent baseband signal. Similarly, $U_s$, $\theta_s$ and the variables related to them are time-invariant during a symbol transmission period, we thus neglect the time index for simplicity. Let us consider a linear time-varying wireless channel $h(\tau, t) = \sum_{i} a_i (t) \delta \left[ \tau - \tau_i (t) \right]$ having the $i$-th path gain $a_i (t)$ and delay $\tau_i(t)$. Then the baseband signal $x_b (t)$ is connected to the transmitted baseband signal $s_b(t)$ in the form of $x_b (t) = \sqrt{A_e} \sum_{i} a_i (t) e^{- \jmath 2 \pi f_c \tau_i (t)} s_b \left[ t - \tau_i (t) \right]$. Upon sampling $x_b (t)$ at multiples of $1/W$, we formulate the sampled outputs as follows \cite{Tse2005Fundamentals} 
	\begin{align}
		\label{eq:SampledCommBasebandRF}
		x_b (m) 
		&= \sqrt{A_e} \sum_{\ell} h_{\ell}(m) s_b (m - \ell),_{}
	\end{align}
	where we have $h_{\ell}(m) 
	= \sum_{i} a_i \left(\frac{m}{W}\right) e^{- \jmath 2 \pi f_c {\tau _i}\left( \frac{m}{W} \right) } \text{sinc}$ $\left[ \ell - \tau_i \left( \frac{m}{W} \right) \right]$ with $\ell = m-n$.

	Upon replacing $x_b(m)$ in \eqref{eq:SampledOutput} by $s_b$ based on \eqref{eq:SampledCommBasebandRF}, we arrive at the equivalent baseband signal model of RAQR-aidded wireless communications in the form of
	\begin{align}
		\label{eq:SampledOutput1}
		\tilde{V}_b (m) 
		&= \sqrt{\varrho } \Phi \sum_{\ell} h_{\ell}(m) s_b (m - \ell),  
	\end{align}
	where $h_{\ell}(m)$ is shown as a complex Gaussian random variable exhibiting the circular symmetry property, namely, $h_{\ell}(m) \sim \mathcal{CN}(0, \sigma_{\ell}^{2})$ \cite{Tse2005Fundamentals}. 
	Furthermore, upon incorporating the additive noise (denoted by $w(m)$) to \eqref{eq:SampledOutput1}, we formulate the noisy discrete-time baseband signal model as follows
	\begin{align}
		\label{eq:SampledOutputPlusNoise}
		\tilde{V}_b (m) = \sqrt{\varrho} \Phi \sum_{\ell} h_{\ell}(m) s_b (m - \ell) + w(m).
	\end{align}
	Even if \eqref{eq:SampledOutputPlusNoise} is capable of characterizing the wideband systems, we emphasize that RAQRs have a narrow instantaneous bandwidth of $\le 10$ MHz \cite{holloway2019detecting}, where only a small frequency range around the resonant frequency can be closely coupled. Based on this situation, we provide the following equivalent baseband signal model for narrowband systems 
	\begin{align}
		\label{eq:NarrowbandSampledOutputPlusNoise}
		\tilde{V}_b (m) = \sqrt{\varrho} \Phi h_{\text{com}}(m) s_b (m) + w(m),
	\end{align}
	where we have $h_{\text{com}}(m) = h_{\ell = 0}(m) \sim \mathcal{CN}(0, \sigma_{\ell = 0}^{2})$.

	\subsubsection{Equivalent Baseband Signal Model for Sensing} 
	The sensing signal $x(t)$ received by the RAQR is an echo (a decay and delayed replica) of the transmitted signal $s(t)$. It has an equivalent baseband signal expressed as $s_b(t) = \sqrt{\frac{A_s c \epsilon_0}{2}} U_s(t - pT) e^{\jmath 2 \pi f_c (t - pT)}$, where $p$ and $T$ are the index of the $p$-th pulse and the pulse-repetition interval, respectively. Let us consider line-of-sight propagation having a path loss of $a(t)$. Then the equivalent baseband signal of $x(t)$ is given by $x_b(t) = a(t) s_b(t - t_d)$, where $t_d \approx \frac{2}{c} (R_0 - vpT)$ is the round-trip time ($R_0$ is the nominal range). Upon sampling this echo at a time instant $m$, we formulate the sampled output as \cite{richards2005fundamentals} 
	\begin{align}
		\label{eq:SampledSensingBasebandRF}
		x_b (m) 
		&= \sqrt{A_e} a(m) s_b \left(m - \frac{2(R_0 - vpT)}{c} \right), 
	\end{align}
	where $v$ is the velocity of the moving target.

	Upon replacing $x_b(m)$ in \eqref{eq:SampledOutput} by $s_b$ based on \eqref{eq:SampledSensingBasebandRF}, we have the equivalent baseband signal model of RAQR-aided wireless sensing in the form of 
	\begin{align}
		\label{eq:SensingSampledOutput}
		\tilde{V}_b (m) 
		&= \sqrt{\varrho} \Phi h_{\text{sen}}(m) s_b \left(m - \frac{ 2 \left(R_0 - vpT \right)}{c} \right), 
	\end{align}
	where $h_{\text{sen}}(m) = a(m)$. We assume that $U_s \left( t - pT - t_d \right) \approx U_s (t - pT)$, implying that the amplitude of $s_b$ is constant over the round-trip period of $t_d = \frac{2(R_0 - vpT)}{c}$. Then we have $s_b \left(m - \frac{2(R_0 - vpT)}{c} \right) \approx e^{- \jmath \frac{4 \pi f_c}{c} (R_0 - vpT)} s_b(m)$, where the phase shift $e^{- \jmath \frac{4 \pi f_c}{c} (R_0 - vpT)}$ can be integrated into $h_{\text{sen}}(m)$. Therefore, we have the noisy form of \eqref{eq:SensingSampledOutput} as follows 
	\begin{align}
		\label{eq:SensingSampledOutput1}
		\tilde{V}_b (m) 
		&= \sqrt{\varrho } \Phi \bar{h}_{\text{sen}}(m) s_b \left(m \right) + w(m), 
	\end{align}
	where $\bar{h}_{\text{sen}}(m) = a(m) e^{- \jmath \frac{4 \pi f_c}{c} (R_0 - vpT)}$.

	\textit{\textbf{Remark 5:}
	\ding{172}
	As seen from both \eqref{eq:NarrowbandSampledOutputPlusNoise} and \eqref{eq:SensingSampledOutput1} for communications and sensing, the RAQR imposes a gain $\varrho$ and a phase shift $\Phi$ to the RF signal. These parameters are determined by both the atomic response and the photodetection scheme selected. Particularly, the phase shift may be integrated into $h(m) \in \{ h_{\text{com}}(m), h_{\text{sen}}(m), \bar{h}_{\text{sen}}(m) \}$ to form an equivalent wireless channel $\Phi h(m)$. For communication purposes, upon exploiting the circular symmetry property, we have $\Phi h(m) \sim \mathcal{CN}(0, \sigma_{\ell}^{2})$ for any given $\Phi \in \{ \Phi_1, \Phi_2 \}$.}
	
	\textit{\ding{173}
	We note that the equivalent baseband signal models of the holistic transceiver (both \eqref{eq:NarrowbandSampledOutputPlusNoise} and \eqref{eq:SensingSampledOutput1}) for communications and sensing are irrelevant to the receiver aperture area $A_e$.}

	\vspace{-1em}
	\subsection{Noise Sources in RAQR-Aided Wireless Systems} 
	In this part, we focus our attention on the main noise sources that affect the RAQR both extrinsically and intrinsically \cite{Fancher2021Rydberg,Tu2024Approaching}, as shown in Fig. \ref{fig:RAQR_and_Classic}(a). 

	\subsubsection{BBR-Induced Extrinsic Noise} 
	\label{subsub:ENS}
	
	The BBR is a source of thermal noise that influences the total coherence time (or dephasing rate) of Rydberg atoms \cite{gallagher1994,Fancher2021Rydberg,beterov2009quasiclassical}. Let us first denote the total dephasing rate by $\varGamma_{2}$, which is obtained as follows \cite{Tu2024Approaching,Fancher2021Rydberg}
	\begin{align}
		\label{eq:CoherenceTime}
		\varGamma_{2} = \gamma_{\text{nat}} + \gamma_{\text{bbr}} + \gamma_{\text{others}}, 
	\end{align}
	where $\gamma_{\text{nat}}$ represents the natural dephasing rate of Rydberg states; furthermore, $\gamma_{\text{bbr}}$ denotes the dephasing rate caused by the BBR. Specifically, for cold atoms, $\gamma_{\text{nat}}$ can be estimated from the natural lifetime of the corresponding Rydberg state $T_{\text{nat}} = \tau_0 n^u$ through $\gamma_{\text{nat}} = \frac{1}{T_{\text{nat}}}$, where $n$ is the principal quantum number, while the values of $\tau_0$ and $u$ can be seen from TABLE 1 of \cite{Fancher2021Rydberg}. Furthermore, $\gamma_{\text{bbr}}$ can be estimated through $\gamma_{\text{bbr}} = \frac{1}{T_{\text{bbr}}}$, where $T_{\text{bbr}} = \frac{ 3 h n^2 }{ 4 \alpha^3 k_B T_{\text{room}} }$ \cite{gallagher1994,Fancher2021Rydberg}. A comprehensive evaluation of $\gamma_{\text{nat}} + \gamma_{\text{bbr}}$ can be found in \cite{beterov2009quasiclassical} for both cold and hot atoms, with their reference values seen in TABLE VIII of \cite{beterov2009quasiclassical}. In \eqref{eq:CoherenceTime}, $\gamma_{\text{others}} = \gamma_0 + \gamma_t + \gamma_r$ accounts for the dephasing rates of other effects \cite{Tu2024Approaching}. Specifically, $\gamma_0$ represents the single-atom Rydberg-EIT dephasing rate caused by finite laser linewidth, laser noise, etc. Furthermore, $\gamma_t$ is the dephasing rate caused by the effect of thermal atoms passing through the beam; finally $\gamma_r$ accounts for Rydberg dipole-dipole interactions \cite{xu2024fast}.

	\subsubsection{Quantum Projection Noise (QPN)}
	It is also known as the atomic shot noise, which is inevitably produced due to the probabilistic collapse of the wavefunction during its measurements. Briefly, when measuring with $N$ quantum-mechanically uncorrelated atoms, the probabilistic collapse from a superposition state to either eigenstate of each atom wavefunction results in the limit of phase measurement, given by $\varphi = \frac{1}{\sqrt{N}}$ \cite{kitching2011atomic,fan2015atom,cox2018quantum}. 
	This leads to the SQL $\frac{U_{x,\text{SQL}}}{\sqrt{\text{Hz}}} = \frac{\hbar}{\mu_{34} \sqrt{N T_2}}$ \cite{Tu2024Approaching}, where $T_2 = \frac{1}{\varGamma_{2}}$ is the coherence time of the EIT process. The SQL will impact the probe beam in the form of a noise, which is finally imposed on the baseband signal. According to \eqref{eq:PhotodetectorOutput_Varying}, the baseband QPN (after LNA) across bandwidth $B$ is derived with the following power 
	\begin{align}
		\nonumber
		N_{\text{QPN}} 
		&= \left[ 2 \sqrt{G} \alpha \widetilde{ \mathcal{P} }
		\kappa (\Omega_l) \cos {\varphi ({\Omega _l})}  \frac{U_{x,\text{SQL}}}{\sqrt{\text{Hz}}} \sqrt{B} \right]^{2} \\
		\label{eq:powSQL}
		&= \varrho c \epsilon_0 \cos^{2} {\varphi ({\Omega _l})} \left( \frac{U_{x,\text{SQL}}}{\sqrt{\text{Hz}}} \right)^{2} B.
	\end{align}

	\begin{figure}[t!]
		\centering
		\includegraphics[width=0.46\textwidth]{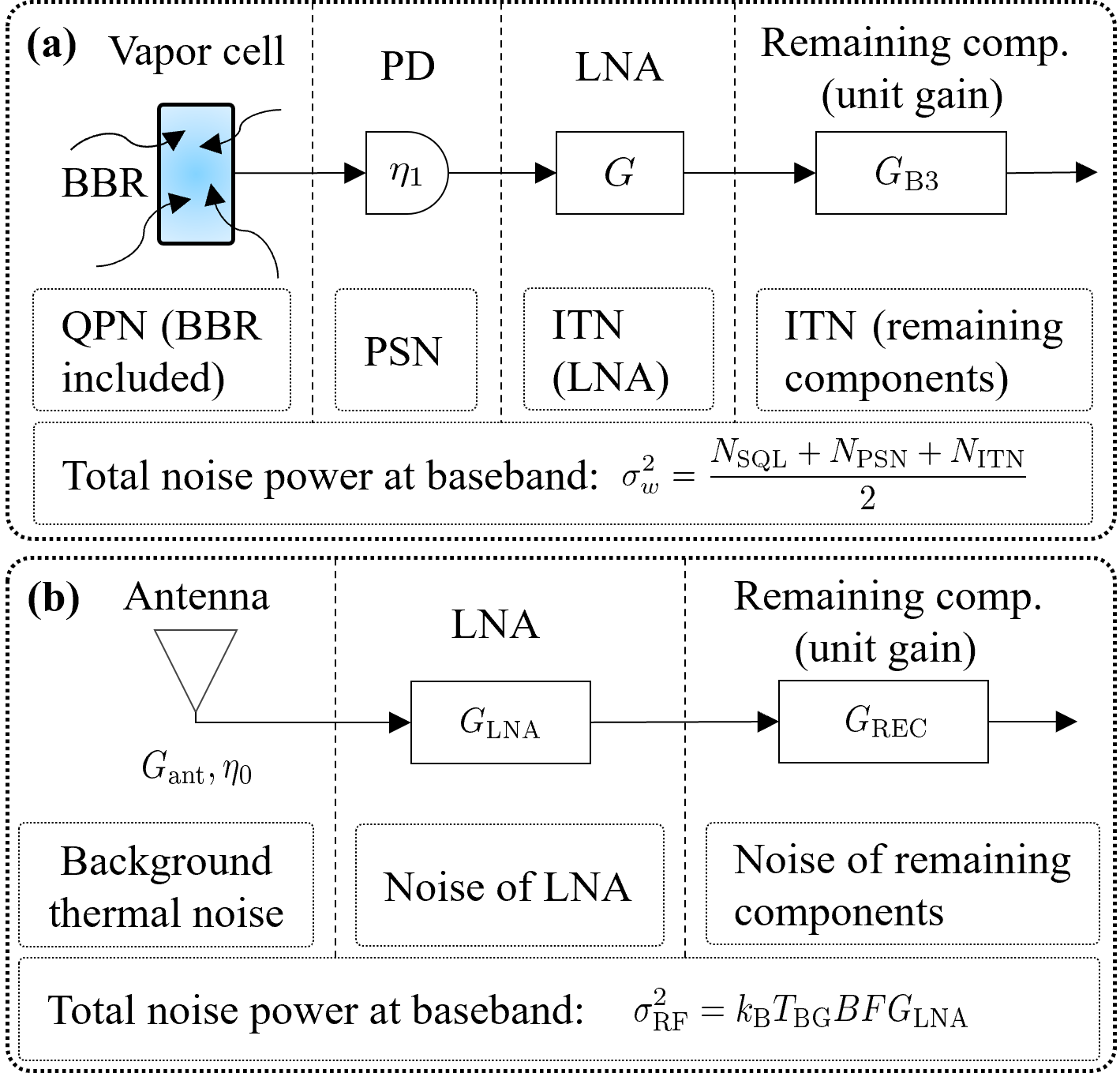}
		\caption{Illustration of main noise sources: (a) RAQRs and (b) classical RF receivers.}
		\vspace{-1.5em}
		\label{fig:RAQR_and_Classic}
	\end{figure}

	\subsubsection{Photon Shot Noise (PSN)} 
	Consider a PIN photodiode. Then two shot noise sources arise in the photodetector, namely, the PSN and the dark-current shot noise. Since the latter is relatively small, we thus focus our attention on the PSN only. For the DIOD scheme, upon denoting the average photocurrent by $\bar{I}_{\text{ph}}$, where $\bar{I}_{\text{ph}} = \alpha \mathcal{P}_{1} (\Omega_{l})$, we formulate the noise power of the PSN (after LNA) across the bandwidth $B$ as \cite{Kasap2013Optoelectronics} 
	\begin{align}
		\label{eq:ShotNoiseDIOD}
		N_{\text{PSN}}^{\text{(D)}}
		= 2qB \bar{I}_{\text{ph}} G. 
	\end{align}
	For the BCOD scheme, let us denote the average photocurrents generated by the two photodetectors by $\bar{I}_{\text{ph}}^{\text{(1)}}$ and $\bar{I}_{\text{ph}}^{\text{(2)}}$, respectively. They correspond to $P_1$ and $P_2$, respectively, and are formulated as $\bar{I}_{\text{ph}}^{\text{(1)}} = \frac{\alpha}{2} [ \mathcal{P}_{l} + \mathcal{P}_{1} (\Omega_{l}) ]$ and $\bar{I}_{\text{ph}}^{\text{(2)}} = \frac{\alpha}{2} [ \mathcal{P}_{l} + \mathcal{P}_{1} (\Omega_{l}) ]$, respectively. Therefore, we formulate the noise power of the PSN (after LNA) across bandwidth $B$ as follows \cite{haus2012electromagnetic,Kasap2013Optoelectronics} 
	\begin{align} 
		\label{eq:ShotNoiseBCOD}
		N_{\text{PSN}}^{\text{(B)}}
		= 2qB \left( \bar{I}_{\text{ph}}^{\text{(1)}} + \bar{I}_{\text{ph}}^{\text{(2)}} \right) G. 
	\end{align}

	\subsubsection{Intrinsic Thermal Noise (ITN)} 
	It is also known as the Johnson noise and it is induced by the random motions of conduction electrons. The ITN power across bandwidth $B$ is 
	\begin{align}
		\label{eq:ThermalNoise}
		N_{\text{ITN}}
		= k_{\text{B}} T B G. 
	\end{align}
	Note that the ITN produced in the system comes from the LNA, HR, and the remaining electronic components (assume $G_{\text{B3}} = 1$ and $T_{\text{B3}}$). Therefore, $T$ becomes an equivalent noise temperature characterizing the combined effect of all these components, namely $T = T_{\text{LNA}} + \frac{1}{G} T_{\text{B3}}$. Based on \cite{orfanidis2002electromagnetic}, 
	the noise temperature of the LNA dominates, yielding $T \approx T_{\text{LNA}}$.

	\subsubsection{Total Noise Power}
	We note that the BBR impacts finally appear as a portion of the QPN. 
	Therefore, we model the baseband noise $w(m)$ in \eqref{eq:SampledOutputPlusNoise}, \eqref{eq:NarrowbandSampledOutputPlusNoise} and \eqref{eq:SensingSampledOutput1} as a complex additive white Gaussian noise process obeying $w(m) \sim \mathcal{CN} (0, \sigma_w^2)$, where we have 
	\begin{align}
		\label{eq:NoisePowerDensity}
		\sigma_w^2 = \frac{N_{\text{QPN}} + N_{\text{PSN}} + N_{\text{ITN}}}{2}. 
	\end{align}
	We further provide the theoretical rationale of \eqref{eq:NoisePowerDensity} as follows
	\begin{itemize}
		\item 
		QPN arises from the probabilistic nature of quantum-state measurements of atoms.  Given a large atomic ensemble, the number of atoms found in a given state obeys a binomial distribution, which approaches a Gaussian one when the ensemble size is large. As seen in \ref{subsec:SQL}, the number of atoms can be obtained as $N = 1.1099 \times 10^8$, which is large enough to ensure the Gaussian distribution. 
		
		\item 
		PSN comes from the Poisson statistics of the random arrival time of individual photons at the photodetector. When the number of detected photons is large, this Poisson process can be well approximated by the Gaussian distribution. Given the probe beam power ${\cal P}_{0} = 20.7$ \textmu W and the energy of one photon $2\pi \hbar \nu_{p}$ with $\nu_{p} = c/\lambda_{p}$, the number of photons per second is obtained as $\frac{{\cal P}_{0}}{2\pi \hbar \nu_{p}} = 8.8722 \times 10^{13}$, also ensuring the Gaussian distribution. 
	\end{itemize}
	Since these noise sources are statistically independent and the ITN is also Gaussian, $w(m)$ obeys the Gaussian distribution. 
	
	\textit{\textbf{Remark 6:}
		\ding{172}
		Different from the signal-independent characteristic of the noise in conventional RF receivers, the noise of RAQRs is signal-dependent, which needs to be taken into account in RAQR-aided wireless designs.
		}
		
		\textit{\ding{173} 
		For typical RAQRs, these noise sources inevitably exist, but the difference is that the noise proportion may be different in different RAQR schemes, such as the DIOD and BCOD schemes. As detailed in Section \ref{sec:PerformanceAnalysis}, different dominant noise contributions correspond to different performance regimes, e.g., SQL and PSL regimes. 
		Notably, the PSL is achievable by the BCOD scheme, where PSN dominates and ITN can be effectively suppressed by exploiting a strong local optical source. However, due to the optical readout scheme employed in RAQRs, it is challenging to achieve the SQL by reducing the PSN from the perspective of noise suppression. A promising research direction may be to develop new schemes harnessing the many-body effect or squeezed states to approach or even surpass the SQL \cite{gong2024RAQRs}.}

	\section{Performance Analysis and Comparison}
	\label{sec:PerformanceAnalysis}

	In this section, we study the received SNR of a narrowband RAQR-aided wireless system. One can further derive the capacity (spectral efficiency) and/or study the bit-error-rate of an RAQR-aided communication system based on the received SNR.  Relying on \eqref{eq:NarrowbandSampledOutputPlusNoise}, \eqref{eq:SensingSampledOutput1}, and \eqref{eq:NoisePowerDensity}, we formulate the SNR of the receive signal as $\mathsf{SNR} = \frac{| \varrho \Phi h(m)|^{2}}{\sigma_w^2} \mathcal{P}_{s}$, which is 
	\begin{align}
		\nonumber
		&\mathsf{SNR} 
		= \frac{ 2 \varrho \cos^2 {\varphi ({\Omega _l})} |{h }(m){|^2}{{\cal P}_{{s}}} }{ \left[ \varrho c \epsilon_0 \cos^{2} {\varphi ({\Omega _l})} \left(  \frac{U_{x,\text{SQL}}}{\sqrt{\text{Hz}}} \right)^{2} + 2qG \alpha {\cal P} + {k_{\rm{B}}} T G \right] B } \\
		\label{eq:NarrowbandReceiveSNR}
		&\xrightarrow[]{\text{maximum}} 
		\frac{ 2 \varrho |{h }(m){|^2}{{\cal P}_{{s}}} }{ \left[ \varrho c \epsilon_0 \left(  \frac{U_{x,\text{SQL}}}{\sqrt{\text{Hz}}} \right)^{2} + 2qG \alpha {\cal P} + {k_{\rm{B}}} T G \right] B }, 
	\end{align}
	where $h(m) \in \{ h_{\text{com}}(m), h_{\text{sen}}(m) \}$ and ${\cal P} \in \{ {{\cal P}_1}({\Omega _l}), {\cal P}_{l} + {{\cal P}_1}({\Omega _l}) \}$ corresponding to the DIOD and BCOD, respectively. The maximum is achieved when $\cos^2 {\varphi ({\Omega _l})} = 1$. This condition always holds for the DIOD scheme because of $\cos^2 {\varphi_1({\Omega _l})} = 1$. For the BCOD scheme, one may configure the phase of the local optical beam $\phi_l$ to retain $\varphi_2 (\Omega_l) = 0$, $\pm \pi$, so that $\cos^2 {\varphi_2({\Omega _l})} = 1$. The imperfect phase alignment of $\varphi_2 (\Omega_l)$ is demonstrated in Section \ref{subsec:impairments}.
	
	According to \eqref{eq:NarrowbandReceiveSNR}, we compare DIOD and BCOD schemes based on their SNR ratio ${\mathsf{Ratio}} = {\mathsf{SNR}}_{\rm{BCOD}} / {\mathsf{SNR}}_{\rm{DIOD}}$. Specifically, we obtain this ratio as follows 
	\begin{align}
		\nonumber
		{\mathsf{Ratio}} 
		&= \frac{{c \epsilon_0 {{\left( {\frac{{{U_{x,{\rm{SQL}}}}}}{{\sqrt {{\rm{Hz}}} }}} \right)}^2} + \frac{{c \epsilon_0 q}}{{2\alpha {{\cal P}_1}({\Omega _l})\kappa _1^2({\Omega _l})}} + \frac{{{k_{\rm{B}}}T}}{{{\varrho _{{\rm{DIOD}}}}}}}}{{c \epsilon_0 {{\left( {\frac{{{U_{x,{\rm{SQL}}}}}}{{\sqrt {{\rm{Hz}}} }}} \right)}^2} + \frac{{c \epsilon_0 q\left[ {1 + \frac{{{{\cal P}_1}({\Omega _l})}}{{{{\cal P}_l}}}} \right]}}{{2\alpha {{\cal P}_1}({\Omega _l})\kappa _2^2({\Omega _l})}} + \frac{{{k_{\rm{B}}}T}}{{{\varrho _{{\rm{BCOD}}}}}}}} \\
		\label{eq:eq:NarrowbandReceiveSNR_Ratio}
		&\xrightarrow[]{ {\frac{{{{\cal P}_1}({\Omega _l})}}{{{{\cal P}_l}}} \to 0} } 
		{\mathsf{Ratio}} \ge 1, 
	\end{align}
	where ${\mathsf{Ratio}} \ge 1$ is due to ${\kappa _1}({\Omega _l}) \le {\kappa _2}({\Omega _l})$, ${\varrho _{{\rm{DIOD}}}} < {\varrho _{{\rm{BCOD}}}}$, and $\mathcal{P}_{l} \gg \mathcal{P}_{1}({\Omega _l})$, suggesting that the BCOD scheme always outperforms the DIOD scheme in terms of the received SNR.

	To gain deeper insights, we investigate the following two specific regimes, which help us understanding the fundamental limit and gaining insights for practical optimizations.

	\vspace{-1em}
	\subsection{SQL Regime}
	\label{subsec:SQL}
	In the SQL regime, only the QPN remains and other noises are perfectly eliminated. This reveals the ultimate fundamental limit of RAQRs with respect to their quantum property. The SNR in this regime becomes
	\begin{align}
		\label{eq:SNR_SQL}
		\mathsf{SNR}_{\text{SQL}} 
		&= 2 Z_0 \left( \frac{ \mu_{34} }{ \hbar } \right)^{2} N T_2 \frac{ |{h }(m){|^2}{{\cal P}_{{s}}} }{ B },
	\end{align}
	which is the identical for both the DIOD and BCOD schemes. 
	It is explicit to observe from \eqref{eq:SNR_SQL} that larger atom number $N$ and long coherence time $T_2$ facilitate higher SNR. 
	
	We further note that $N$ is determined by a combination of the atomic density $N_0$, the optical volume containing Rydberg atoms $V$, and the population rate $\varUpsilon$ to the Rydberg state, namely $N = \varUpsilon N_0 V$. Assuming that the laser beams are Gaussian beams, we approximate the receiver volume, where the probe/coupling beams are spatially-overlapped in the vapor cell, as a cylinder. Upon assuming that the FWHMs of the probe and coupling beams are identical, and the radius of the cylinder is given by $r_0 = F_p/\sqrt{2 \ln{2}}$, and the volume containing Rydberg atoms is $V = A_p d$, where $A_p = \pi r_0^2 = \pi F_p^2 / (2 \ln{2})$ is the cross-sectional area of the cylinder.

	Further substituting $N = \varUpsilon N_0 V$, $V = \pi F_p^2 d / (2 \ln{2})$, and $T_2 = \frac{1}{\varGamma_{2}}$ into \eqref{eq:SNR_SQL}, we arrive at 
	\begin{align}
		\label{eq:SNR1_SQL}
		\mathsf{SNR}_{\text{SQL}} 
		&= \frac{ \pi Z_0 {F_p^2} }{ \ln{2} } \left( \frac{ \mu_{34} }{ \hbar } \right)^{2} \left( \frac{ \bar{N}_0 d }{ \varGamma_{2} } \right) \frac{ |{h }(m){|^2}{{\cal P}_{{s}}} }{B},
	\end{align}
	where $\bar{N}_0 \triangleq \varUpsilon N_0$ is defined as the effective atomic density. 
	\textit{\textbf{Remark 7:}
	In the SQL regime, the SNR grows linearly with the effective atomic density $\bar{N}_0$ and the length of the vapor cell $d$, while grows quadratically with the FWHM of the probe beam. More particularly, the SNR theoretically grows unbounded as these parameters increase infinitely. In practice, $\bar{N}_0$ may be restricted due to the so-called blockade sphere, where only one atom can be excited to Rydberg state within the region of the blockade sphere.}

	\vspace{-1em}
	\subsection{PSL Regime}
	In this regime, only the PSN is considered. This reveals the fundamental limit of employing different photodetection schemes. 
	This PSL is achievable for the BCOD scheme because the ITN can be made relatively low compared to the PSN by increasing the power of the local optical source ${\cal P}_{l}$, as seen in \eqref{eq:NarrowbandReceiveSNR}. By contrast, this regime only holds for the DIOD scheme when the detected power of the probe beam ${{\cal P}_1}({\Omega _l})$ is high so that the PSN becomes higher than that of the ITN. Despite this, the PSL regime result of the DIOD scheme is still meaningful. Specifically, we reformulate the received SNR as 
	\begin{align}
		\nonumber
		\hspace{-0.3em} \mathsf{SNR}_{\text{PSL}} 
		&= \frac{4 \alpha Z_0}{ q } 
		\begin{cases}
			\mathcal{P}_{1} (\Omega_{l}) \kappa_1^2 (\Omega_l) \frac{ |{h }(m){|^2}{{\cal P}_{{s}}} }{B}, &\text{DIOD},\\
			\frac{ \mathcal{P}_{l} \mathcal{P}_{1} (\Omega_{l}) }{ {\cal P}_{l} + {{\cal P}_1}({\Omega _l}) } \kappa_2^2 (\Omega_l) \frac{ |{h }(m){|^2}{{\cal P}_{{s}}} }{B},  &\text{BCOD},
		\end{cases} \\
		\nonumber
		&\approx \frac{4 \alpha Z_0}{ q } 
		\begin{cases}
			\mathcal{P}_{1} (\Omega_{l}) \kappa_1^2 (\Omega_l) \frac{ |{h }(m){|^2}{{\cal P}_{{s}}} }{B}, &\text{DIOD},\\
			\mathcal{P}_{1} (\Omega_{l}) \kappa_2^2 (\Omega_l) \frac{ |{h }(m){|^2}{{\cal P}_{{s}}} }{B},  &\text{BCOD},
		\end{cases} \\
		\label{eq:SNR_PSL}
		&= \frac{{\eta_1 A_p}}{{2 \pi \hbar f_p }} 
		\begin{cases}
			U_p^2 (\Omega_{l}) \kappa_1^2 (\Omega_l) \frac{ |{h }(m){|^2}{{\cal P}_{{s}}} }{B}, &\text{DIOD},\\
			U_p^2 (\Omega_{l}) \kappa_2^2 (\Omega_l) \frac{ |{h }(m){|^2}{{\cal P}_{{s}}} }{B},  &\text{BCOD},
		\end{cases} 
	\end{align}
	where the approximation is obtained by using $\mathcal{P}_{l} \gg \mathcal{P}_{1}({\Omega _l})$.  
	
	\textit{\textbf{Remark 8:}
	\ding{172}
	As seen from \eqref{eq:SNR_PSL}, given bandwidth $B$ and channel ${h}(m)$, $\mathsf{SNR}_{\text{PSL}}$ is determined by the product of $U_p^2 (\Omega_{l})$ and $\kappa^2 (\Omega_l)$, $\kappa (\Omega_l) \in \{ \kappa_1 (\Omega_l), \kappa_2 (\Omega_l) \}$ instead of determined by only the output probe beam. Therefore, to achieve higher received SNR, one may (jointly) optimize the power of the output probe beam, the detuning frequency, and LO signal strength to obtain an optimal product of $U_p^2 (\Omega_{l}) \kappa^2 (\Omega_l)$.}
		
	\textit{\ding{173}
	We further note that $\kappa_1 (\Omega_l) = \kappa_2 (\Omega_l)$ only when $\Delta_{p,c,l} = 0$ without any detuning. This implies that both the DIOD and BCOD schemes achieve the same SNR in the PSL regime, as numerically verified in Fig. \ref{fig:ParameterOptimization}(a)(b) of Section \ref{subsec:OP}. Otherwise, for the optimized $\Delta_{p,c,l}$, the received SNR of the BCOD scheme is higher than that of the DIOD scheme, as numerically verified in Fig. \ref{fig:sim_result_1} of Section \ref{subsec:performance}.}

	\vspace{-1em}
	\subsection{Comparison to Classical RF Receivers}
	
	\subsubsection{Performance Comparison}
	We present a typical single-antenna reception scheme for classical RF receivers in Fig. \ref{fig:RAQR_and_Classic}(b). The system consists of an antenna (gain $G_{\text{ant}}$ and radiation efficiency $\eta_0$), an LNA (gain $G_{\text{LNA}}$), and the remaining components (equivalent gain $G_{\text{REC}}$), such as mixers, lowpass filters and ADCs. Without loss of generality, we assume $G_{\text{REC}} = 1$. Upon considering narrowband systems, we derive the equivalent baseband input-output signal model as 
	\begin{align}
		\label{eq:RFNarrowbandSampledOutputPlusNoise}
		\tilde{V}_{0} (m) = \sqrt{ A_{\text{iso}} \varrho_{0} } h(m) s_b (m) + w_{0}(m),
	\end{align}
	where $\varrho_{0} = \eta_0 G_{\text{ant}} G_{\text{LNA}}$ represents the gain of the classical RF receiver, $A_{\text{iso}} = \lambda^2 / (4 \pi)$ is the effective receiver aperture of an isotropic antenna, and $w_{0}$ is the AWGN of the RF receiver. Specifically, the noise includes the thermal noise due to the background thermal noise, the LNA noise, and the noise of the remaining components, namely we have $w(m) \sim \mathcal{CN} (0, \sigma_{\text{RF}}^{2} )$, where $\sigma_{\text{RF}}^{2} = k_{\text{B}} T_{\text{BG}} B F G_{\text{LNA}}$, $T_{\text{BG}}$ is the temperature of the background (assume room temperature of $T_{\text{BG}} = 300$ K), $F$ represents the noise factor of the holistic system.  Here $\sigma_{\text{RF}}^{2}$ can also be expressed in decibel as $- 174 {\rm{ dBm/Hz}} + 10\log B + NF + G_{\text{LNA}}$, where $NF = 10 \log(F)$ represents the system noise figure. For example, $NF$ is $6$ dB and $9$ dB for the base station (BS) and the user equipment (UE), respectively, at the frequency range of 5G FR1 n104 \cite{3GPP_IMT}. Based on these discussions, we have 
	\begin{align}
		\label{eq:RFNarrowbandReceiveSNR}
		\mathsf{SNR}_{0} 
		&= \frac{{{A_{{\rm{iso}}}}\varrho _0 |{h }(m){|^2}{{\cal P}_{{s}}}}}{ \sigma_{\text{RF}}^{2} } 
		= \frac{ \eta_0 {A_{{\rm{iso}}}} G_{\text{ant}} |{h }(m){|^2}{{\cal P}_{{s}}} }{ k_{\text{B}} T_{\text{BG}} B F }. 
	\end{align}

	We compare RAQRs and classical RF receivers in terms of the SNR ratio, where the SQL and PSL regimes are characterized as follows. 
	Based on \eqref{eq:SNR_SQL} and \eqref{eq:RFNarrowbandReceiveSNR}, we formulate $\mathsf{Ratio}_{0} = \mathsf{SNR}_{\text{SQL}} / \mathsf{SNR}_{0}$ in the SQL regime as  
	\begin{align}
		\label{eq:RAQR_RF_SQL}
		\mathsf{Ratio}_{0} 
		= 2 Z_0 \left( \frac{ {\mu _{34}} }{ {\hbar} } \right)^{2} \left( \frac{ {{\bar N}_0} }{ {\varGamma_2} } \right) \left( \frac{ {A_p} d }{ A_{\rm{iso}} } \right) \left( \frac{{k_{\rm{B}}}{T_{{\rm{BG}}}}F}{{\eta _0}{G_{{\rm{ant}}}}} \right).
	\end{align}
	Additonally, based on \eqref{eq:SNR_PSL} and \eqref{eq:RFNarrowbandReceiveSNR}, we formulate $\overline{\mathsf{Ratio}}_{0} = \mathsf{SNR}_{\text{PSL}} / \mathsf{SNR}_{0}$ in the PSL regime as 
	\begin{align}
		\label{eq:RAQR_RF_PSL}
		\overline{\mathsf{Ratio}}_{0} 
		= \frac{ \eta_1 {U_p^2}({\Omega _l}) {\kappa ^2}({\Omega _l}) }{ 2 \pi \hbar f_p } 
		\left( \frac{{A_p}}{A_{{\rm{iso}}}} \right) \left( \frac{{k_{\rm{B}}}{T_{{\rm{BG}}}}F}{{\eta _0}{G_{{\rm{ant}}}}} \right).
	\end{align}

	
	\begin{table*}[!t]
	\renewcommand{\arraystretch}{1.06}
	\caption{\textsc{Configuration of RAQR parameters in simulations.}}
	\label{tab:parameters}
	\centering
	\tabcolsep = 0.16cm
	\resizebox{\linewidth}{!}{
		\begin{tabular}{|l|l|l|}
			\hline 
			\rowcolor{cyan!10} 
			\multicolumn{3}{|c|}{ \textbf{Electron transitions} } \\
			\hline 
			\rowcolor{cyan!5}
			\textbf{Parameter} & \textbf{Value} & \textbf{Unit} \\
			\hline Vapor cell length &  $d=10$  & cm \\
			\hline Atomic density &  $N_{0}=4.89 \times 10^{10}$  &  $\text{cm}^{-3}$  \\
			\hline Population rate &  $\varUpsilon=1 \%$  &  /  \\
			\hline Dipole moment of $\ket{1}$ \textrightarrow $\ket{2}$  &  $\mu_{12}=2.2327 q a_{0}$  &  C/m  \\
			\hline Dipole moment of  $\ket{2}$ \textrightarrow $\ket{3}$  &  $\mu_{23}=0.0226 q a_{0}$  &  C/m  \\
			\hline Dipole moment of  $\ket{3}$ \textrightarrow $\ket{4}$  &  $\mu_{34}=1443.45 q a_{\mathrm{0}}$  &  C/m  \\
			\hline Decay rate of  $\ket{2}$  &  $\gamma_{2}=5.2$  & MHz \\
			\hline Decay rate of  $\ket{3}$  &  $\gamma_{3}=3.9$  & kHz \\
			\hline Decay rate of  $\ket{4}$  &  $\gamma_{4}=1.7$  & kHz \\
			\hline Total dephasing rate &  $\varGamma_2=5$  & MHz \\
			\hline Coherence time &  $T_2=0.2$  & \textmu s \\
			\hline
		\end{tabular}
		\begin{tabular}{|l|l|l|}
			\hline 
			\rowcolor{cyan!10} 
			\multicolumn{3}{|c|}{ \textbf{Laser beams and RF signals} } \\
			\hline 
			\rowcolor{cyan!5} 
			\textbf{Parameter} & \textbf{Value} & \textbf{Unit} \\
			\hline Probe beam wavelength &  $\lambda_{p}=852$  & nm \\
			\hline Coupling beam wavelength &  $\lambda_{c}=510$  & nm \\
			\hline Probe beam power &  $\mathcal{P}_{0}=20.7$  & \textmu W \\
			\hline Coupling beam power &  $\mathcal{P}_{c}=17$  & mW \\
			\hline Local optical beam power &  $\mathcal{P}_{l}=30$  & mW \\
			\hline Probe/coupling beam radius &  $r_{0}=1.7$  & mm \\
			\hline LO signal amplitude &  $U_{y}=0.0661$  & V/m  \\
			\hline Carrier frequency &  $f_{c}=6.9458$  & GHz \\
			\hline \begin{tabular}{l} 
				Frequency difference \\
				between LO and RF
			\end{tabular} &  $f_{\delta}=150$  & kHz \\
			\hline RF bandwidth &  $B=100$  & kHz \\
			\hline
		\end{tabular}
		\begin{tabular}{|l|l|l|}
			\hline 
			\rowcolor{cyan!10} 
			\multicolumn{3}{|c|}{ \textbf{Antenna, LNA, and others} } \\
			\hline 
			\rowcolor{cyan!5} 
			\textbf{Parameter} & \textbf{Value} & \textbf{Unit} \\
			\hline Antenna efficiency &  $\eta_0 = 0.7$  & / \\
			\hline BS antenna element gain (5G FR1 n104) &  $G_{\text{ant}}=5.5$  & dB \\
			\hline UE antenna element gain (5G FR1 n104) &  $G_{\text{ant}}=0$  & dB \\
			\hline System noise figure of BS and UE &  $NF=6, 9$  & dB \\
			\hline LNA gain of classical RF receiver &  $G_{\text{LNA}}=60$  & dB \\
			\hline Noise temperature of LNA &  $T_{\text{LNA}}=100$  & K \\
			\hline Room temperature &  $T_{\text{room}}=300$  & K \\
			\hline Quantum efficiency &  $\eta_1=0.8$  &  /  \\
			\hline Photodetector LNA gain &  $G=30$  & dB \\
			\hline Photodetector LNA noise temperature &  $T=100$  & K \\
			\hline Photodetector load resistance &  $R=1$  & Ohm \\
			\hline
		\end{tabular}}
		\vspace{-1.5em}
	\end{table*}

	\subsubsection{Implementation-Oriented Comparison}
	Built upon mature semiconductor technologies, conventional RF receivers exhibit compact integration, low cost, and wide bandwidth (typically MHz–GHz), with their performance limited by thermal noise. By contrast, RAQRs exhibit significant differences and trade-offs in their architecture and implementation. Specifically, RAQRs exploit multi-level atomic coherence and precisely controlled laser excitation, offering intrinsically self-calibrated field measurements and continuous tunability across an exceptionally broad frequency range (DC-THz). However, these benefits may come at the cost of increased system complexity due to the tight specifications of stable optical, thermal, and frequency-locking subsystems, which may presently limit compactness, scalability, and bandwidth (typically kHz–MHz). Compared to conventional RF receivers, RAQRs are still in the infancy of their engineering applications.

	\vspace{-0.8em}
	\section{Simulation Results}
	\label{sec:Simulations}
	
	Upon following the configurations in \cite{jing2020atomic}, we consider a four-level electron transition scheme 6S\textsubscript{\scalebox{0.8}{1/2}} \textrightarrow 6P\textsubscript{\scalebox{0.8}{3/2}} \textrightarrow 47D\textsubscript{\scalebox{0.8}{5/2}} \textrightarrow 48P\textsubscript{\scalebox{0.8}{3/2}}. The corresponding parameters are presented in TABLE \ref{tab:parameters}. We note that the parameters of the antenna and LNAs obey those in \cite{balanis2016antenna,belostotski2020down,santamaria2022comparison,3GPP_IMT}. We note that the received RF signal of RAQR may be a consequence of the signal propagated through various channel conditions while satisfying the narrowband condition due to the narrowband receiver property of RAQRs. Without loss of generality, we consider free space propagation having a large-scale fading of $-30 + 10 \beta \log(1/L)$, where the distance is $L = 1500$ m and the pathloss exponent is $\beta = 2.0$. The amplitude of the received RF signal is thus $-71.8$ dBV/m. 
	Unless otherwise stated, our simulations will follow the above configurations. We will first validate the correctness of our model, then showcase a simple parameter optimization, and further simulate the effects of several critical parameters on the received SNRs \eqref{eq:SNR1_SQL}, \eqref{eq:SNR_PSL}, SNR ratios \eqref{eq:RAQR_RF_SQL}, \eqref{eq:RAQR_RF_PSL}, and the receiver sensitivity \eqref{eq:sensitivity}. Finally, we investigate the influence of several practical impairments on the performance of RAQRs.

	\vspace{-1em}
	\subsection{Validation of the Proposed Signal Model}
	\label{subsec:ModelValidation}
\begin{figure*}[t!]
		\centering
		\subfloat[]{
			\includegraphics[width=0.296\textwidth]{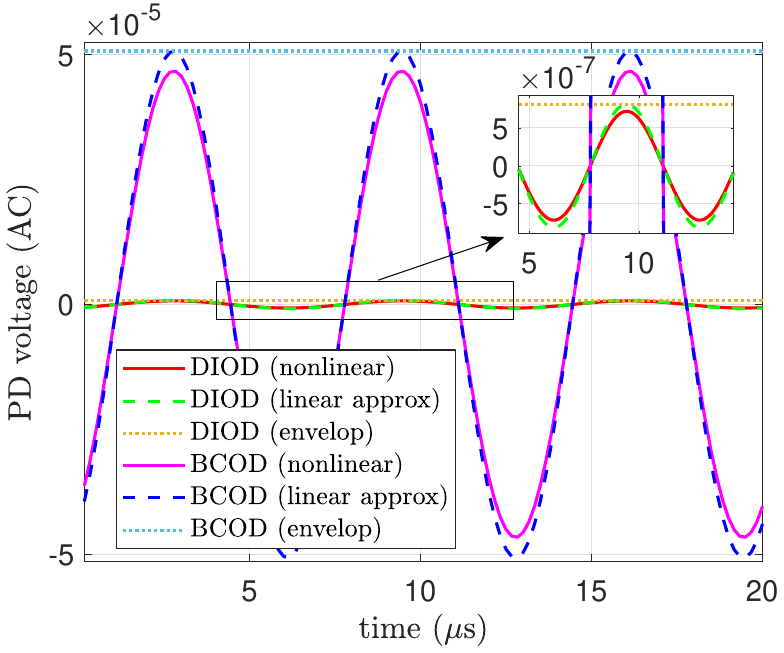}} 
		\subfloat[]{
			\includegraphics[width=0.3\textwidth]{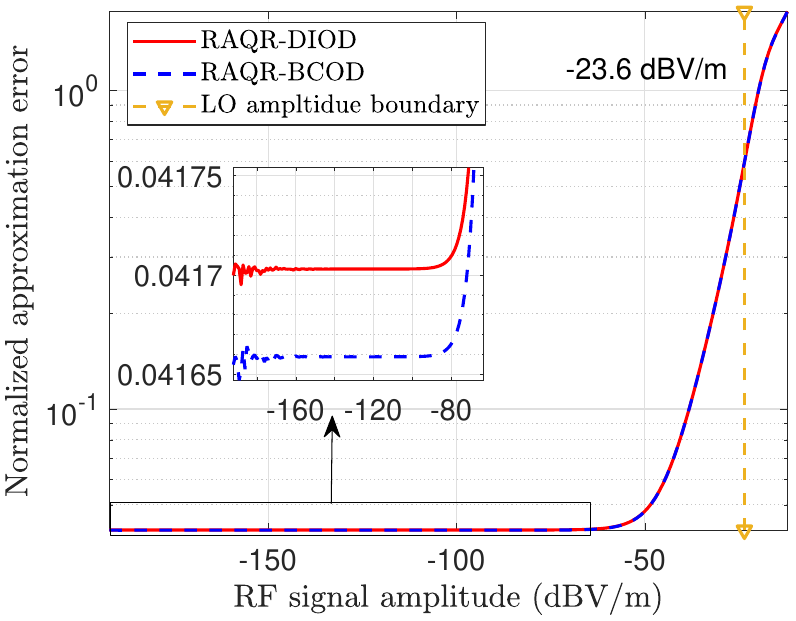}} 
		\subfloat[]{
			\includegraphics[width=0.302\textwidth]{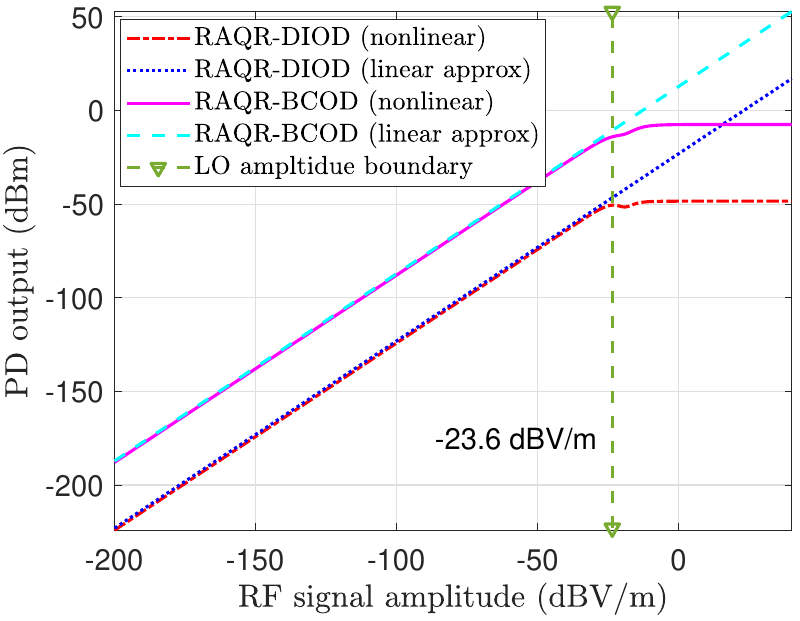}}
		\vspace{-0.5em}
		\caption{(a) Waveform, (b) normalized approximation error, and (c) input-output transfer function characterized by our models.}
		\vspace{-1.2em}
		\label{fig:Validation_waveform_error}
	\end{figure*}

	We validate our signal model by showing the coincidence of waveforms between the AC component in \eqref{eq:PhotodetectorOutput_Varying} constructed by our model and the exact AC obtained by removing the DC of $V^{\text{(D)}}(t) 
	= \sqrt{G} I_{\text{ph}}^{\text{(D)}} (t)$ and $V^{\text{(B)}}(t) 
	= \sqrt{G} I_{\text{ph}}^{\text{(B)}} (t)$ in \eqref{eq:PhotodetectorOutputDIOD} and \eqref{eq:PhotodetectorOutputBCOD}, respectively. We select an observation window of $0.02$ ms to capture the waveforms, during which we keep the amplitude of the RF signal constant. 
	Then, we quantify the normalized approximation error between the approximated AC and the exact AC. 
	Finally, we compare the input-output transfer function depicted by our models to exact ones. 
	For all above simulations, we consider $\Delta_{p,c,l} = 0$ without any scanning. Their results obtained are shown in Fig. \ref{fig:Validation_waveform_error}.

	\begin{figure*}[t!]
		\centering
		\subfloat[]{
			\includegraphics[width=0.296\textwidth]{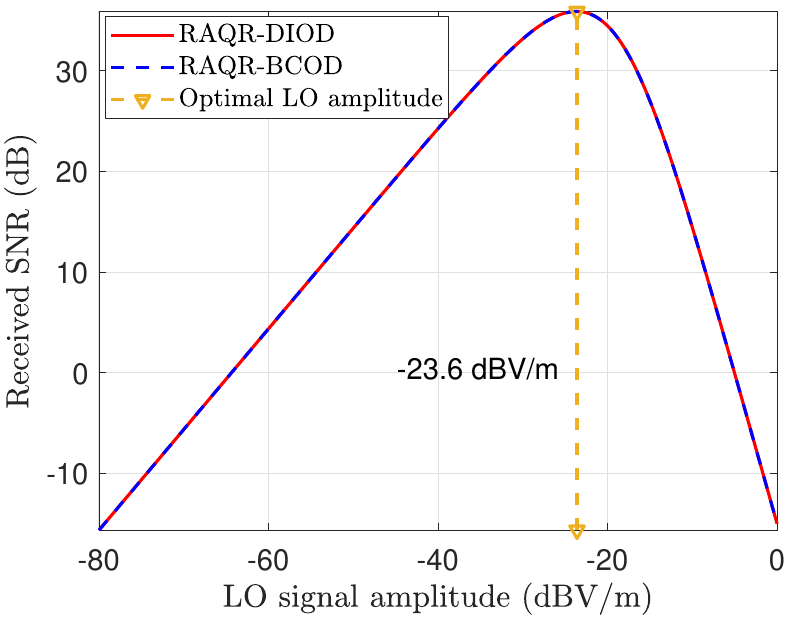}}
		\subfloat[]{
			\includegraphics[width=0.302\textwidth]{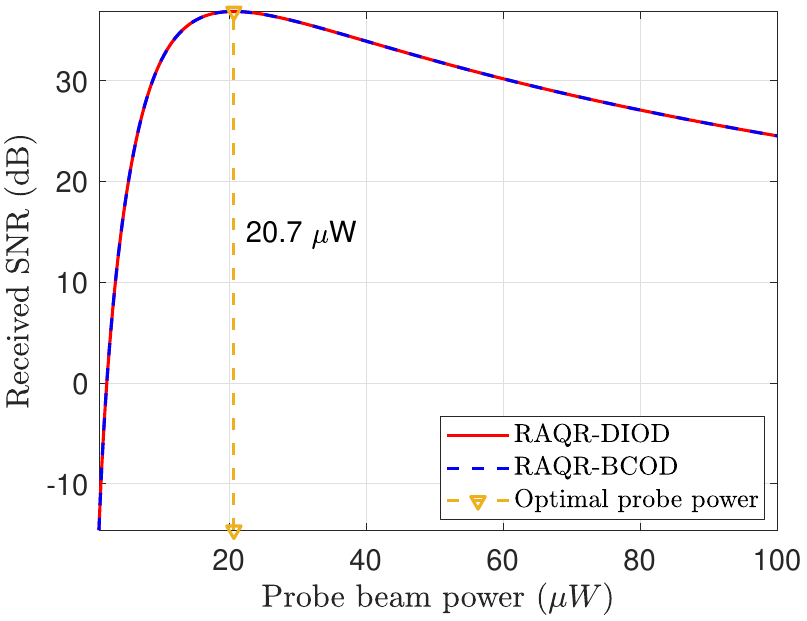}}
		\subfloat[]{
			\includegraphics[width=0.292\textwidth]{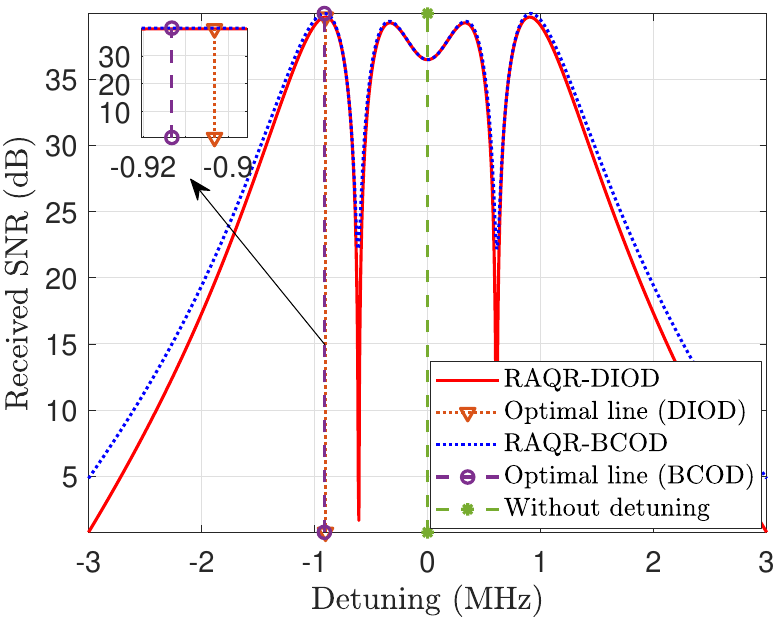}}\\
		\subfloat[]{
			\includegraphics[width=0.296\textwidth]{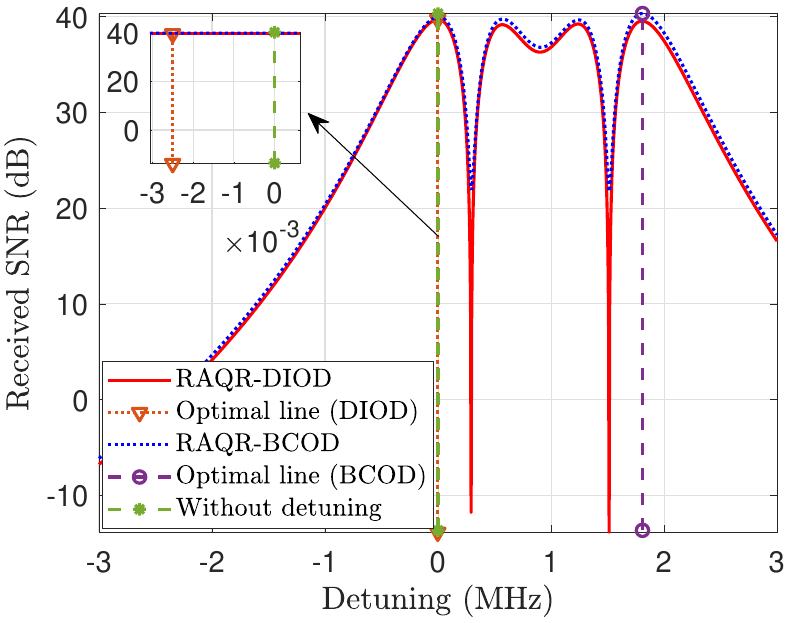}}
		\subfloat[]{
			\includegraphics[width=0.291\textwidth]{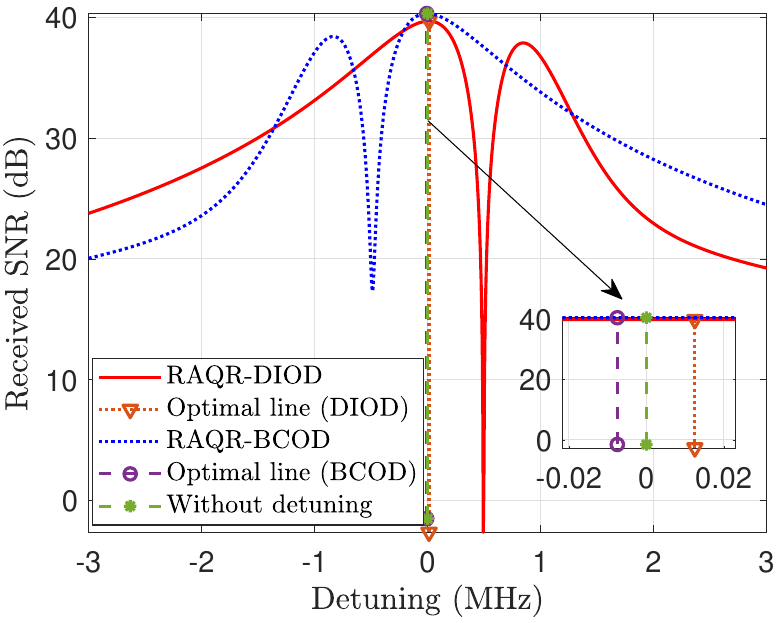}}
		\subfloat[]{
			\includegraphics[width=0.304\textwidth]{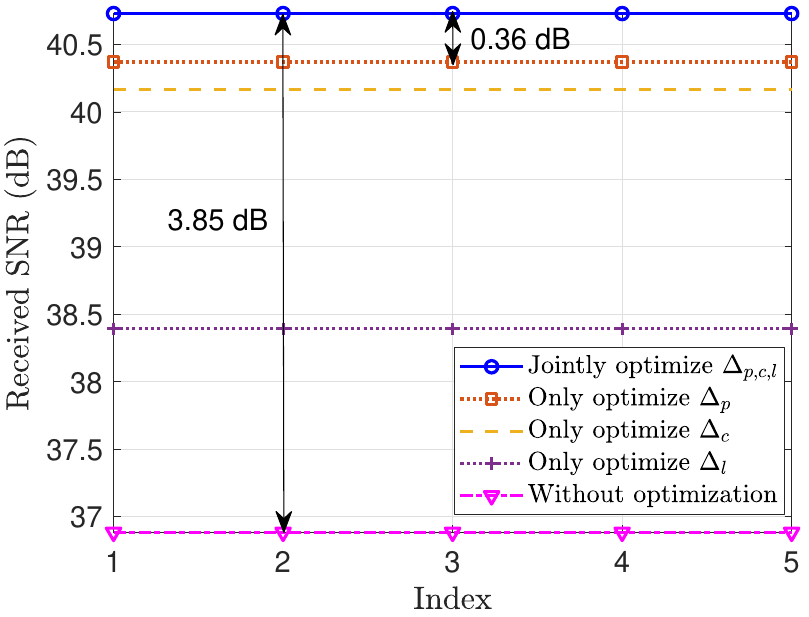}}
		\vspace{-0.5em}
		\caption{Optimization of parameters: (a) LO signal amplitude $U_y$, and (b) probe beam power $\mathcal{P}_{0}$, (c) probe beam detuning $\Delta_{p}$, (d) coupling beam detuning $\Delta_{c}$, and (e) LO detuning $\Delta_{l}$. (f) Compare the joint optimization of $\Delta_{p,c,l}$ to their independent optimizations and to the case without detuning optimization of $\Delta_{p,c,l} = 0$.}
		\vspace{-1.5em}
		\label{fig:ParameterOptimization}
	\end{figure*}

	It is observed from Fig. \ref{fig:Validation_waveform_error}(a) that the waveform characterized by our linear approximation model well coincides with the waveform of the realistic nonlinear RAQR for both DIOD and BCOD schemes. The envelopes characterized by our linear approximation models can accurately trace the exact waveforms. We note that the envelope is constant because we fix the detunings without any scanning. Furthermore, we observe from Fig. \ref{fig:Validation_waveform_error}(b) that the proximity of waveforms can be characterized by the normalized approximation error. When the RF signal's amplitude is small enough compared to LO's amplitude, the errors of DIOD and BCOD schemes are quite small. However, the errors gradually increase when the RF signal's amplitude is close to LO's amplitude. This is because the condition $U_{y} \gg U_{x}$ becomes invalid as the RF signal's amplitude increases compared to a fixed-amplitude LO. Similar phenomenon can be observed from Fig. \ref{fig:Validation_waveform_error}(c). The nonlinear transfer function of RAQRs can be accurately characterized by our linear approximations only requiring $U_{y} \gg U_{x}$. It is noteworthy that the transfer functions have a large linear dynamic range with a properly reconfigured LO, and the input-output curves of Fig. \ref{fig:Validation_waveform_error}(c) are consistent with \cite{jing2020atomic}.

	\vspace{-1em}
	\subsection{Optimization of Parameters}
	\label{subsec:OP}
	
	We then perform simple optimizations to diverse parameters, including the LO signal amplitude, probe beam power, and detuning frequencies of the probe beam, coupling beam, and the LO signal. These optimized parameters facilitate an improvement of the received SNR. We consider the received SNR \eqref{eq:SNR_PSL} in the PSL regime as an example in our simulations. 
	Since these parameters are coupled, their joint optimization is highly intractable. To tackle this problem, we exploit the idea of alternating optimization. Specifically, we first optimize the LO signal amplitude by fixing the probe beam power as $29.8$ \textmu W and $\Delta_{p,c,l} = 0$. Then we sequentially optimize $\mathcal{P}_{0}$, $\Delta_{p}$, $\Delta_{c}$, and $\Delta_{l}$, where the pre-optimized values are substituted into the current process. We optimize one round (only once) for each parameter and acquire their maxima by exhaustive search, where their results are portrayed in Fig. \ref{fig:ParameterOptimization}.

	We observe from Fig. \ref{fig:ParameterOptimization}(a)(b) that the optimal values of the LO amplitude and the probe power, maximizing the received SNR \eqref{eq:SNR_PSL}, are $-23.6$ dBV/m ($0.0661$ V/m) and $20.7$ \textmu W, respectively. Improper configuration may lead to drastically reduction of the received SNR. The trends of both figures are consistent with physics experimental papers \cite{jing2020atomic} and \cite{wu2017enhanced}. 
	Another observation from Fig. \ref{fig:ParameterOptimization}(a)(b) is that the received SNR of the BCOD is similar to that of the DIOD. This is because $\kappa_1 (\Omega_{l}) = \kappa_2 (\Omega_{l})$ due to $\Delta_{p,c,l} = 0$, as discussed in \textbf{Remark 4} of Section \ref{subsec:RAQR} and \textbf{Remark 8} of Section \ref{sec:PerformanceAnalysis}. 
	
	We can see from Fig. \ref{fig:ParameterOptimization}(c)(d)(e) that the commonly used configurations $\Delta_{p,c,l} = 0$ are usually not optimal, as indicated by the line marked as ``Without detuning". Additionally, we observe that the optimal lines indicate the maximal received SNR. The jointly optimized values for the DIOD and BCOD schemes are $\Delta_{p,c,l} = \{ -0.9033, -0.0025, 0.0125 \}$ MHz and $\Delta_{p,c,l} = \{ -0.9133, 1.8090, -0.0075 \}$ MHz, respectively. 
	Furthermore, in Fig. \ref{fig:ParameterOptimization}(f), we showcase the performance improvement of our joint optimization of $\Delta_{p,c,l}$ compared to the independent optimization of $\Delta_{p,c,l}$ employed in \cite{wu2023theoretical, wu2024atomic} and to the case without detuning optimization of $\Delta_{p,c,l} = 0$ (as noted in \textbf{Remark 1}). In this comparison, we employ the optimal LO signal power and the optimal probe beam power obtained in Fig. \ref{fig:ParameterOptimization}(a)(b) for all compared cases. From Fig. \ref{fig:ParameterOptimization}(f), we observe that our joint optimization yields an extra improvement of $0.36$ dB over the best received SNR of the independent optimization, and improves the case without optimization by about $3.8$ dB.

	We elaborate by mentioning that although the additional 0.36 dB improvement attained by joint optimization appears modest, it highlights the conceptual value of parameter co-design and represents the upper bound of the achievable performance. Joint optimization also indicates a potential enhancement in robustness under parameter fluctuations thanks to the increased optimization degrees-of-freedom. The associated computational overhead of joint optimization is minor, as the analytical formulation allows efficient implementation at the cost of limited additional iterations.

	\begin{figure*}[t!]
		\centering
		\subfloat[]{
			\includegraphics[width=0.3\textwidth]{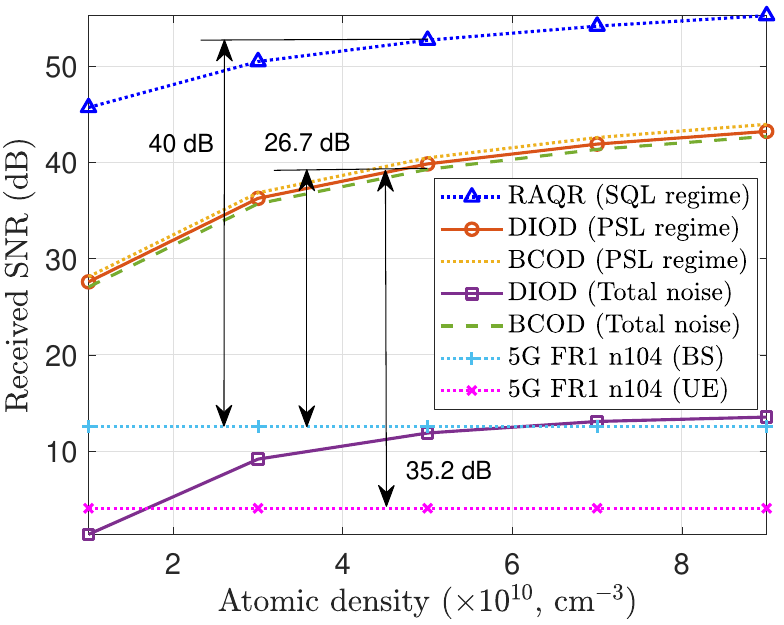}}
		\subfloat[]{
			\includegraphics[width=0.3\textwidth]{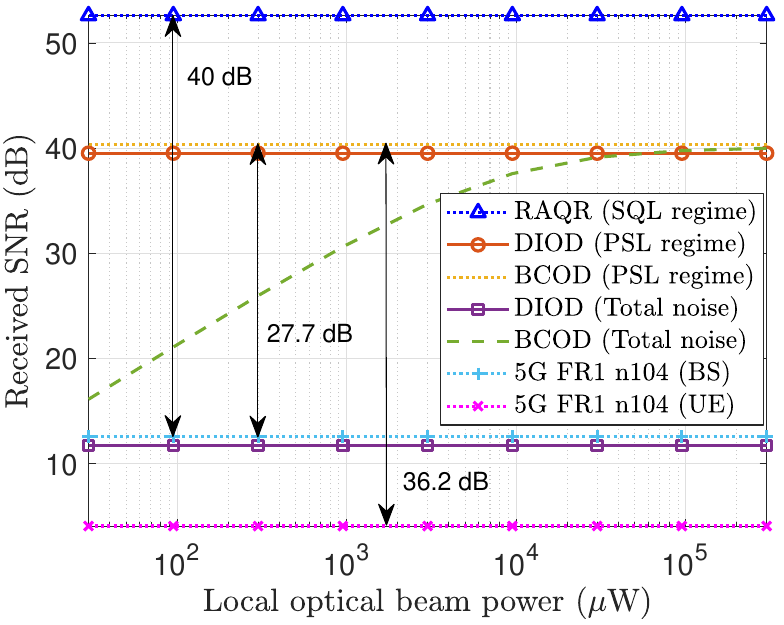}}
		\subfloat[]{
			\includegraphics[width=0.3\textwidth]{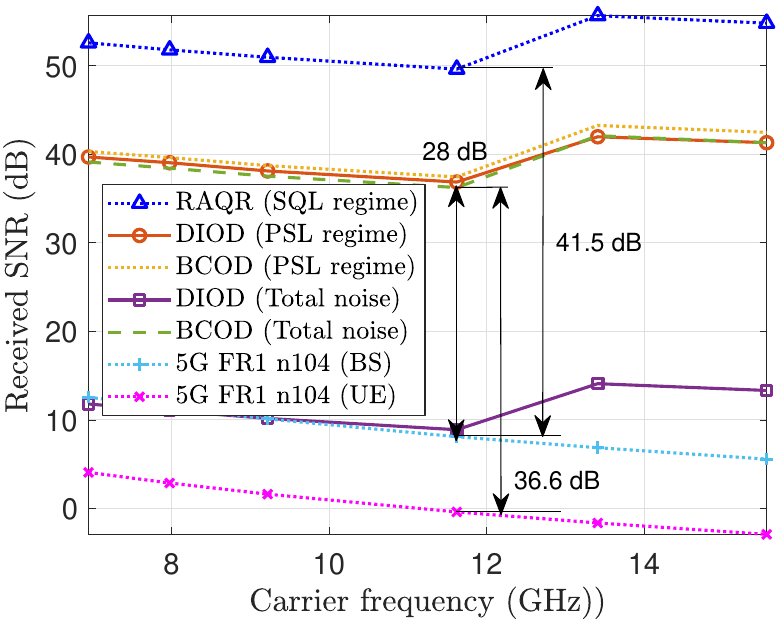}}
		\vspace{-0.5em}
		\caption{The received SNR vs. (a) the atomic density $N_0$, (b) local optical beam power ${\cal P}_{l}$, and (c) carrier frequency $f_c$.}
		\vspace{-1.5em}
		\label{fig:sim_result_1}
	\end{figure*}

	\vspace{-1em}
	\subsection{Performance of the RAQR-Aided Wireless Systems}
	\label{subsec:performance}

	In this section, we presents our simulation results by following the optimized parameters. Since the operating frequency is 6.9458 GHz that is within the frequency range of 5G FR1 n104. Therefore, the classic RF receiver is chosen as the 5G-BS and 5G-UE, respectively, where their parameters are from the 3GPP specification \cite{3GPP_IMT} and are given in TABLE \ref{tab:parameters}. Furthermore, we present several cases of the RAQR, namely
	\begin{itemize}
		\item 
		\textbf{RAQR (SQL regime)} --- Only the QPN is considered and the received SNR is obtained by \eqref{eq:SNR1_SQL}. 
		
		\item 
		\textbf{DIOD (PSL regime)} --- The DIOD is used and only the PSN is considered. The received SNR is given by \eqref{eq:SNR_PSL}. 
		
		\item 
		\textbf{BCOD (PSL regime)} --- The BCOD is used and only the PSN is considered. The received SNR is given by \eqref{eq:SNR_PSL}. 
		
		\item 
		\textbf{DIOD (Total noise)} --- The DIOD is employed and all noises (QPN, PSN, ITN) are considered. The received SNR is obtained by using \eqref{eq:NarrowbandReceiveSNR}. 
		
		\item 
		\textbf{BCOD (Total noise)} --- The BCOD is employed and all noises (QPN, PSN, ITN) are considered. The received SNR is obtained by using \eqref{eq:NarrowbandReceiveSNR}. 
	\end{itemize}

	\begin{table}[!t]
		\small
		\renewcommand{\arraystretch}{1.1}
		\caption{\textsc{Parameter configurations of Cs atoms for receiving different carrier frequencies.}}
		\label{tab:CarrierFrequencyConfig}
		\centering
		\tabcolsep = 0.14cm
		\resizebox{\linewidth}{!}{
			\begin{tabular}{!{\vrule width0.6pt}c|c|c|c|c|c|c|c!{\vrule width0.6pt}}			
				\Xhline{0.6pt}
				\rowcolor{cyan!10} 
				$\ket{3}$ \textrightarrow $\ket{4}$ 
				& $\mu_{34}$
				& $f_c$
				& $U_y$
				& $\mathcal{P}_{0}$
				& $\Delta_{p}$
				& $\Delta_{c}$
				& $\Delta_{l}$ \\
				\Xhline{0.6pt}
				\rowcolor{cyan!5} 
				/
				& Cm
				& GHz
				& V/m
				& \textmu W
				& \multicolumn{3}{c|}{\cellcolor{cyan!5} \tabincell{l}{MHz (DIOD, BCOD)}}\\
				\Xhline{0.6pt}
				\tabincell{c}{47D\textsubscript{\scalebox{0.8}{5/2}} \textrightarrow\\ 48P\textsubscript{\scalebox{0.8}{3/2}}}
				& 1443.4$q a_0$ 
				& 6.9458 
				& 0.0661
				& 20.7 
				& \tabincell{c}{-0.9033,\\-0.9133}
				& \tabincell{c}{-0.0025,\\1.8090}
				& \tabincell{c}{0.0125,\\-0.0075}\\
				\Xhline{0.6pt}
				\tabincell{c}{45D\textsubscript{\scalebox{0.8}{5/2}} \textrightarrow\\ 46P\textsubscript{\scalebox{0.8}{3/2}}}
				& 1316.6$q a_0$ 
				& 7.9752 
				& 0.0708
				& 20.3 
				& \tabincell{c}{-0.8832,\\-0.8932}
				& \tabincell{c}{-0.0025,\\1.7690}
				& \tabincell{c}{0.0125,\\-0.0025}\\
				\Xhline{0.6pt}
				\tabincell{c}{43D\textsubscript{\scalebox{0.8}{5/2}} \textrightarrow\\ 44P\textsubscript{\scalebox{0.8}{3/2}}}
				& 1195.7$q a_0$ 
				& 9.2186 
				& 0.0794
				& 20.6 
				& \tabincell{c}{-0.8982,\\-0.9083}
				& \tabincell{c}{-0.0025,\\1.7990}
				& \tabincell{c}{0.0075,\\-0.0025}\\
				\Xhline{0.6pt}
				\tabincell{c}{40D\textsubscript{\scalebox{0.8}{5/2}} \textrightarrow\\ 41P\textsubscript{\scalebox{0.8}{3/2}}}
				& 1025.1$q a_0$ 
				& 11.6187 
				& 0.0912
				& 20.4 
				& \tabincell{c}{-0.8832,\\-0.8932}
				& \tabincell{c}{-0.0025,\\1.7740}
				& \tabincell{c}{0.0075,\\-0.0075}\\
				\Xhline{0.6pt}
				\tabincell{c}{66S\textsubscript{\scalebox{0.8}{1/2}} \textrightarrow\\ 66P\textsubscript{\scalebox{0.8}{3/2}}}
				& 2055.4$q a_0$ 
				& 13.4078 
				& 0.0501
				& 20.4 
				& \tabincell{c}{-0.9883,\\-0.9883}
				& \tabincell{c}{0.0025,\\1.9291}
				& \tabincell{c}{0.0325,\\-0.0225}\\
				\Xhline{0.6pt}
				\tabincell{c}{63S\textsubscript{\scalebox{0.8}{1/2}} \textrightarrow\\ 63P\textsubscript{\scalebox{0.8}{3/2}}}
				& 1862.7$q a_0$ 
				& 15.5513 
				& 0.0537
				& 20.1 
				& \tabincell{c}{-0.9583,\\-0.9633}
				& \tabincell{c}{0.0025,\\1.8791}
				& \tabincell{c}{0.0225,\\-0.0125}\\
				\Xhline{0.6pt}
			\end{tabular}
		}
		\vspace{-1.5em}
	\end{table}

	\textbf{Received SNR versus (vs.) atomic density ($N_0$):} 
	Upon fixing all other parameters and varying the atomic density with $N_0 = (1 \sim 9) \times 10^{10}$ $\text{cm}^{-3}$, we plot the results in Fig. \ref{fig:sim_result_1}(a). 
	It is observed that the received SNR of RAQR logarithmically increases in dB vs. the atomic density. All curves of the RAQR in both the PSL and SQL regimes exhibit significant SNR gains over the classical 5G-BS and 5G-UE. For example, when $N_0 = 5 \times 10^{10}$ $\text{cm}^{-3}$, the SNR gain over the 5G-BS can be $27$ dB and $40$ dB in the PSL and SQL regime, respectively, and becomes higher when compared to 5G-UE. It is noted from Fig. \ref{fig:sim_result_1}(a) that the ``DIOD (Total noise)" has a huge SNR gap to the ``DIOD (PSL regime)" and does not outperform the 5G-BS. However, this can be substantially improved by the ``BCOD (Total noise)", which can approach the ``BCOD (PSL regime)" and exhibits $26.7$ dB ($35.2$ dB) over the 5G-BS (5G-UE), when the atomic density is $5 \times 10^{10}$ $\text{cm}^{-3}$.

	\textbf{Received SNR vs. local optical beam power (${\cal P}_{l}$):}
	We fix other parameters and vary ${\cal P}_{l}$ from $30$ \textmu W to $300$ mW. This influences the received SNR of ``BCOD (Total noise)", as shown in Fig. \ref{fig:sim_result_1}(b). From the figure that the received SNRs except the ``BCOD (Total noise)" remain constant across the entire range of ${\cal P}_{l}$, once the other parameters are given. The ``BCOD (Total noise)" exhibits a large SNR gap compared to the ``BCOD (PSL regime)" when ${\cal P}_{l}$ is small. The gap becomes narrower as ${\cal P}_{l}$ increases, implying that the ``BCOD (Total noise)" can approach the limit of the PSL regime thanks to the suppression of the ITN due to large ${\cal P}_{l}$.

	\textbf{Received SNR vs. carrier frequency ($f_c$):} 
	It is noteworthy that RAQRs are limited to receiving RF signals at specific discrete frequencies that are coupled to different but specific transitions between two Rydberg states. We follow the configurations of Cs atoms in \cite{simons2016simultaneous} for receiving different carrier frequencies. The transition from the ground state to the excited state is 6S\textsubscript{\scalebox{0.8}{1/2}} \textrightarrow 6P\textsubscript{\scalebox{0.8}{3/2}}, which is realized obeying the same configurations of laser beams as stated above. The carrier frequencies and several optimized parameters are provided in TABLE \ref{tab:CarrierFrequencyConfig}. Based on these configurations, we present our simulation results in Fig. \ref{fig:sim_result_1}(c).

	\begin{figure*}[t!]
		\centering
		\subfloat[]{
			\includegraphics[width=0.3\textwidth]{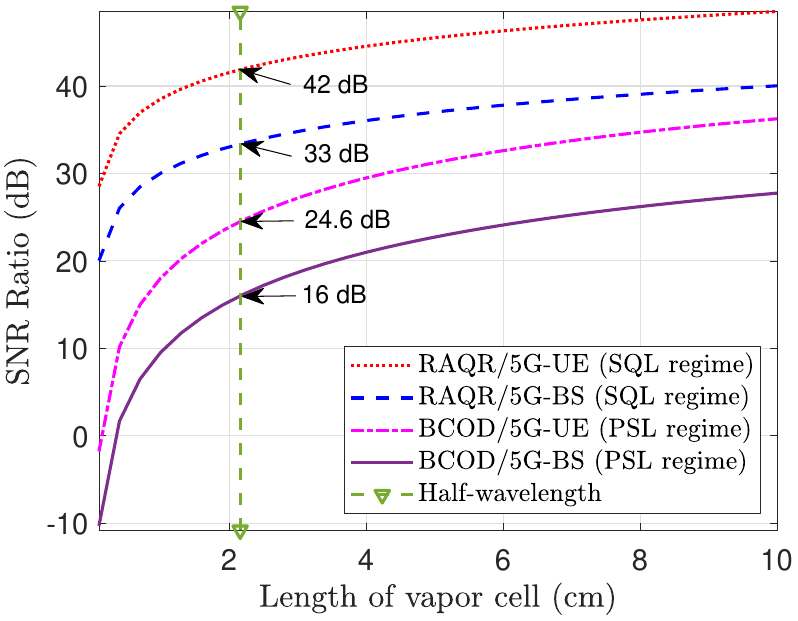}}
		\subfloat[]{
			\includegraphics[width=0.29\textwidth]{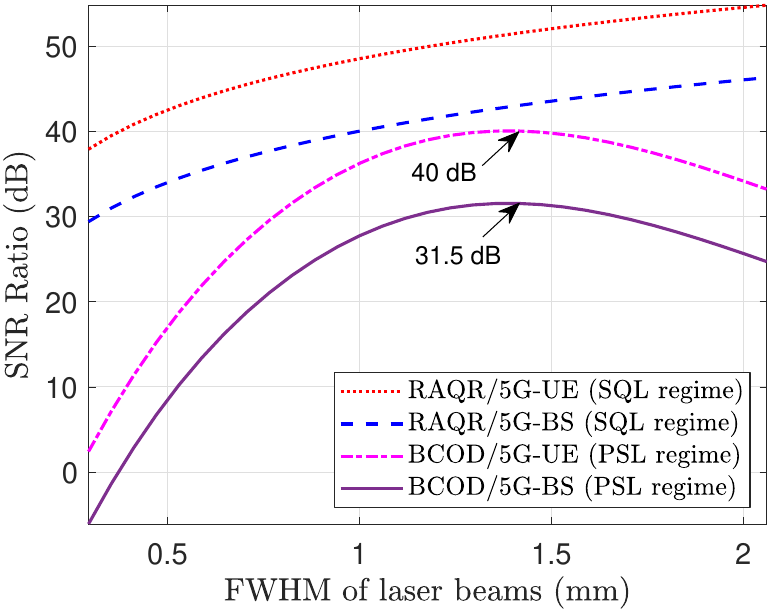}}
		\subfloat[]{
			\includegraphics[width=0.306\textwidth]{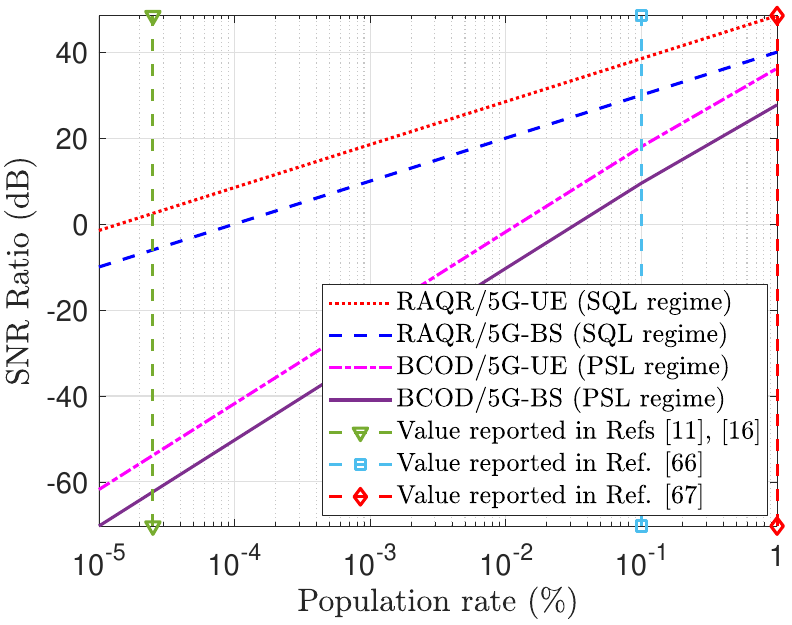}}
		\vspace{-0.5em}
		\caption{The SNR ratio vs. (a) the length of vapor cell $d$, (b) FWHM of laser beams $F_p$, and (c) population rate $\varUpsilon$.}
		\vspace{-1.2em}
		\label{fig:sim_result_2}
	\end{figure*}

	\begin{figure*}[t!]
		\centering
		\subfloat[]{
			\includegraphics[width=0.3\textwidth]{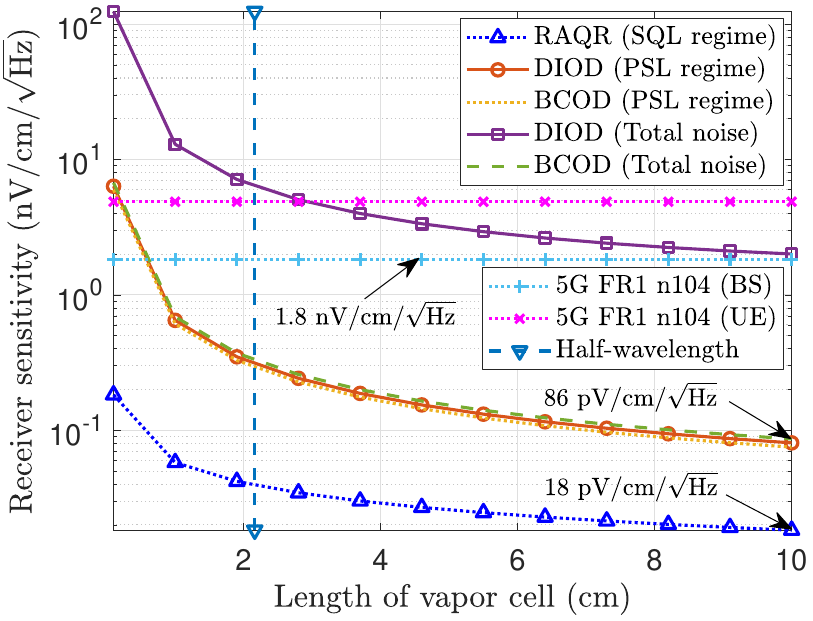}}
		\subfloat[]{
			\includegraphics[width=0.298\textwidth]{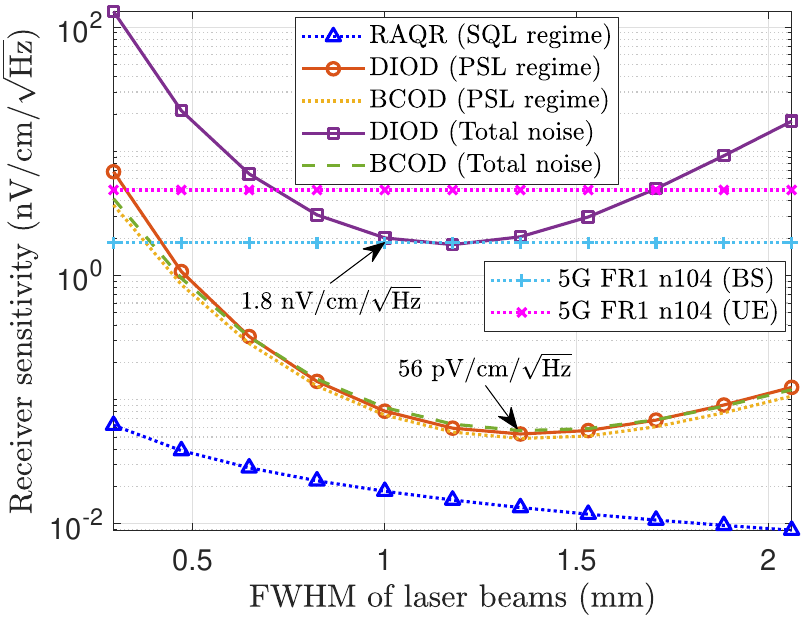}}
		\subfloat[]{
			\includegraphics[width=0.3\textwidth]{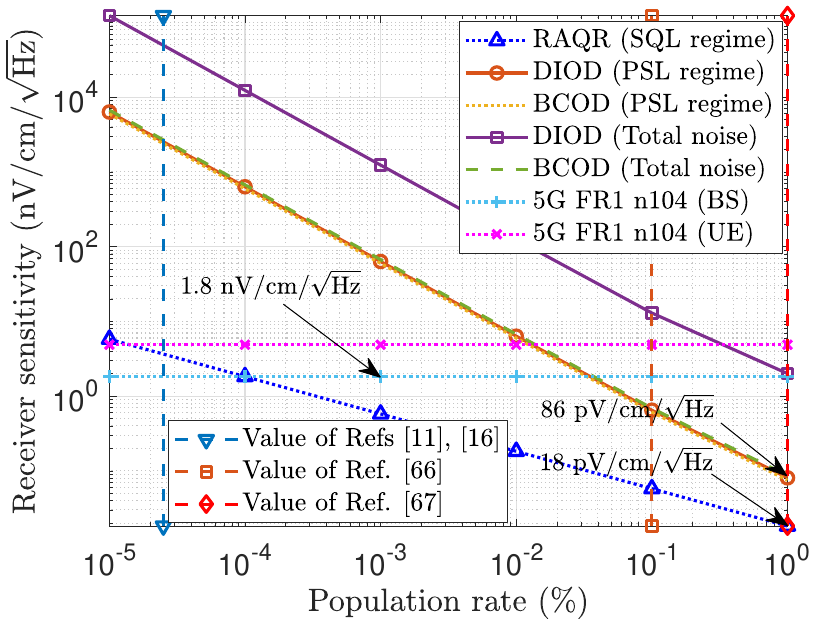}}
		\vspace{-0.5em}
		\caption{The receiver sensitivity vs. (a) the length of vapor cell $d$, (b) FWHM of laser beams $F_p$, and (c) population rate $\varUpsilon$.}
		\vspace{-1.5em}
		\label{fig:sim_result_3}
	\end{figure*}

	Let us observe the RAQR of all configurations in Fig. \ref{fig:sim_result_1}(c), the curves slightly decrease from $6.9458$ GHz to $11.6187$ GHz, then suddenly increase from $11.6187$ GHz to $13.4078$ GHz, and become flat at the rest of the frequency range. The trend is mainly determined by the dipole moment, but is less related to the carrier frequency. The higher the dipole moment, the higher the received SNR.  By contrast, the received SNRs of classical 5G-BS and 5G-UE gradually drop vs. the carrier frequency. This trend is essentially due to the reduction of its effective receiver aperture. The SNR difference between the ``BCOD (Total noise)" and classical 5G-BS (5G-UE) exceeds $28$ dB ($36$ dB) and becomes larger with the increase of the carrier frequency, as seen in Fig. \ref{fig:sim_result_1}(c). This trend holds for the RAQR in the SQL regime, but the SNR differences are further augmented to $41.5$ dB and $50$ dB at the $11.6187$ GHz.

	\textbf{SNR ratio vs. length of vapor cell $d$, FWHM of laser beams $F_p$, and population rate $\varUpsilon$:} 
	We further examine how the SNR ratios \eqref{eq:RAQR_RF_SQL}, \eqref{eq:RAQR_RF_PSL} are influenced by these critical parameters. We only present the result of the BCOD scheme as a benefit of its PSL achievable, as portrayed in Fig. \ref{fig:sim_result_2}.

	Firstly, we observe from Fig. \ref{fig:sim_result_2}(a) that the SNR ratios in both the SQL and PSL regimes gradually grow as the cell length increases and they become higher than $0$ over a range of $0.4 \sim 10$ cm. This reveals that the RAQR outperforms the classical 5G-BS/5G-UE with a consistent increase of the SNR gain. At $6.9458$ GHz, the cell length of the half-wavelength size is also marked in Fig. \ref{fig:sim_result_2}(a), where an extra SNR gain of $16$ dB ($24.6$ dB) can be achieved by the ``BCOD (PSL regime)" over 5G-BS (5G-UE). This SNR gain is further augmented to $33$ dB ($42$ dB) in the SQL regime.

	Then we observe from Fig. \ref{fig:sim_result_2}(b) that in the PSL regime, the SNR ratio has a maximum of $31.5$ dB ($40$ dB) at an optimal FWHM of $\sim 1.4$ mm. Both smaller and larger FWHMs degrade the SNR ratio. This is essentially due to the inherent connection between the FWHM and the probe beam power. By contrast, the SNR ratio in the SQL regime always increases without FWHM restriction.

	Lastly, we examine the critical impact of the population rate $\varUpsilon$ in Fig. \ref{fig:sim_result_2}(c). In both the SQL and PSL regimes, the SNR ratio increases once we enlarge the population rate. To elaborate more on the high impact of $\varUpsilon$, we consider different configurations of $\varUpsilon = 0.01\% \times \frac{1}{400}$ and $\varUpsilon = 0.1\%$ reported in \cite{fan2015atom, Fancher2021Rydberg} and \cite{bohaichuk2022origins}, respectively. They lead to widely different conclusions, where the first configuration reveals the RAQR is inferior to the classical 5G-BS (5G-UE) even in the SQL regime, while the latter configurations illustrate that the RAQR in both regimes outperforms the classical counterparts. Therefore, improving $\varUpsilon$ is the key to ensure the superiority of RAQRs. We also notice from \cite{brown2023very} that $\varUpsilon \rightarrow 1\%$ is achievable, which is what we applied in this article.

	\textbf{Receiver sensitivity vs. length of vapor cell $d$, FWHM of laser beams $F_p$, and population rate $\varUpsilon$:} 
	In addition to the received SNR and the SNR ratio, we also showcase the sensitivity vs. several critical parameters. Upon exploiting the relationship of $\frac{ |{h }(m){|^2}{{\cal P}_{{s}}} }{ B } = \frac{ c \epsilon_0 \left| U_{x} \right|^{2} }{ 2 B } = \frac{ \left| U_{x} \right|^{2} }{ 2 Z_0 B }$, we can obtain the sensitivity of the RAQR in different regimes by setting the received SNRs \eqref{eq:SNR1_SQL}, \eqref{eq:SNR_PSL}, \eqref{eq:NarrowbandReceiveSNR} to $0$ dB, respectively, namely 
	\begin{align}
		\label{eq:sensitivity}
		\hspace{-0.5em} \frac{U_{x}}{\sqrt{\text{Hz}}} = 
		\begin{cases}
			\frac{U_{x,\text{SQL}}}{\sqrt{\text{Hz}}} = \frac{\hbar}{\mu_{34} \sqrt{N T_2}}, &\text{SQL},\\
			\frac{U_{x,\text{PSL}}}{\sqrt{\text{Hz}}} = \frac{ \sqrt{q} }{ \kappa (\Omega_l) \sqrt{ 2 \alpha \mathcal{P}_{1} (\Omega_{l}) } }, &\text{PSL},\\
			\sqrt{ \left(  \frac{U_{x,\text{SQL}}}{\sqrt{\text{Hz}}} \right)^{2} + \left(  \frac{U_{x,\text{PSL}}}{\sqrt{\text{Hz}}} \right)^{2} + \frac{{k_{\rm{B}}} T G Z_0}{\varrho} }, &\text{Total}.
		\end{cases} 
	\end{align}
	Likewise, the sensitivity of the classical RF receivers can be obtained as $\frac{U_{x}}{\sqrt{\text{Hz}}} = \sqrt{ \frac{ 2 Z_0 k_{\text{B}} T_{\text{BG}} F }{ \eta_0 {A_{{\rm{iso}}}} G_{\text{ant}} } }$ by configuring the received SNR \eqref{eq:RFNarrowbandReceiveSNR} to $0$ dB. Based on all above formulas, we present our simulation results of the sensitivity in Fig. \ref{fig:sim_result_3}. 
	
	It is observed that the trend of the sensitivity shown in Fig. \ref{fig:sim_result_3} is contrary to that of the SNR ratio portrayed in Fig. \eqref{fig:sim_result_2}. More particularly, the sensitivity of the classical 5G-BS (5G-UE) is $1.8$ nV/cm/$\sqrt{\text{Hz}}$ ($4.88$ nV/cm/$\sqrt{\text{Hz}}$), while the sensitivity of the RAQR can be as low as $86$ pV/cm/$\sqrt{\text{Hz}}$ and $18$ pV/cm/$\sqrt{\text{Hz}}$ in the PSL and SQL regimes, respectively, exhibiting the extremely high sensitivity of RAQRs for detecting RF signals. We also note that the proposed BCOD scheme is capable of achieving the PSL.

		\begin{figure*}[t!]
			\centering
			\subfloat[]{
				\includegraphics[width=0.306\textwidth]{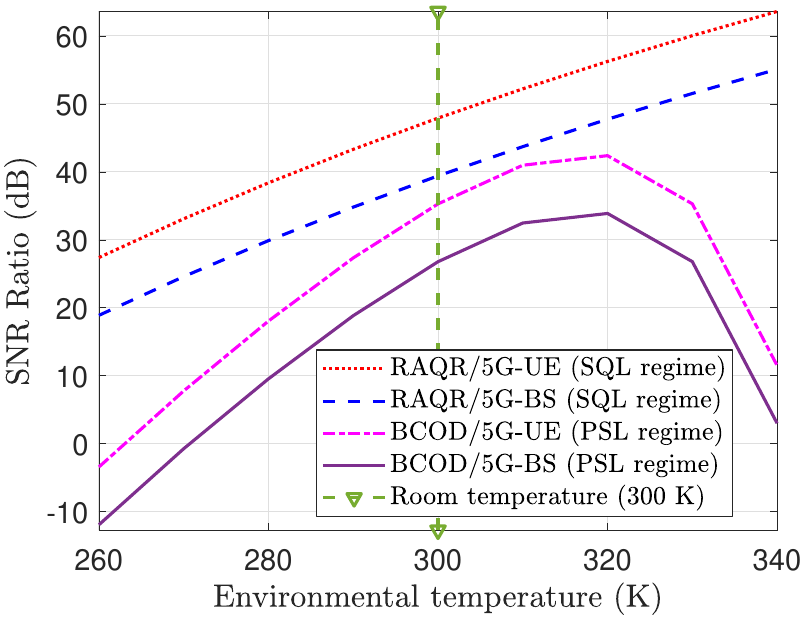}}
			\subfloat[]{
				\includegraphics[width=0.302\textwidth]{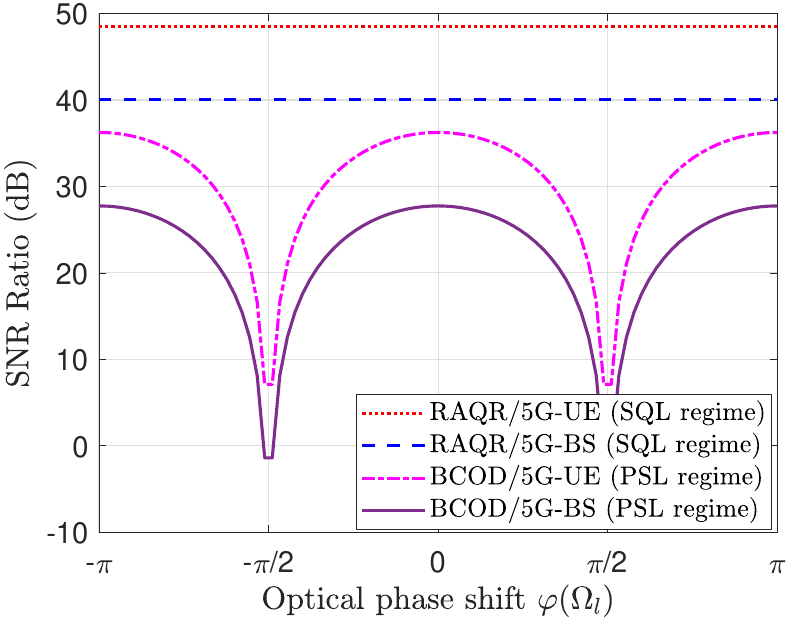}}
			\subfloat[]{
				\includegraphics[width=0.3\textwidth]{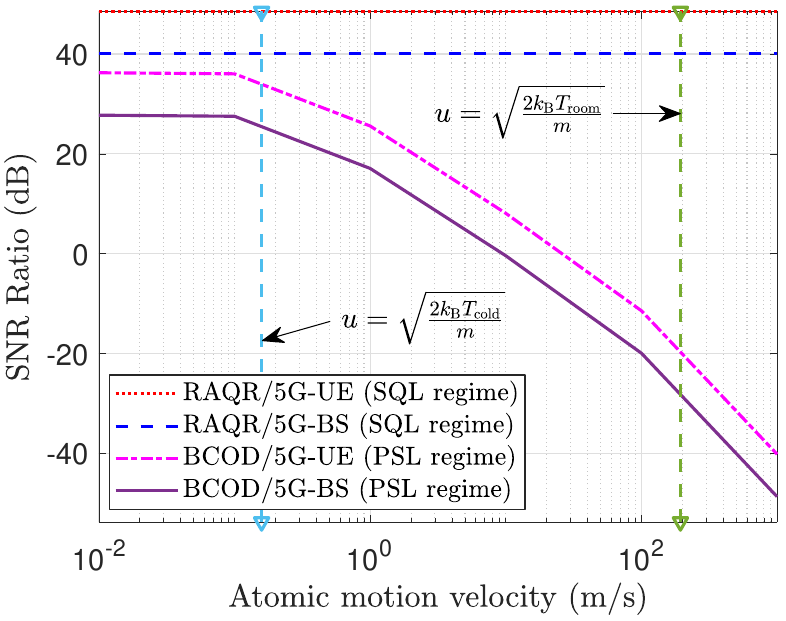}} \\
			\subfloat[]{
				\includegraphics[width=0.304\textwidth]{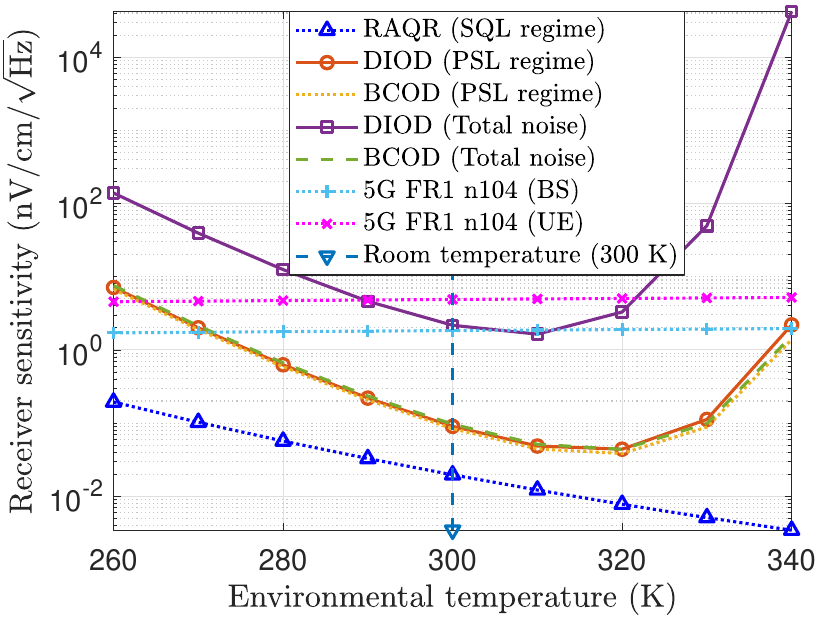}}
			\subfloat[]{
				\includegraphics[width=0.3\textwidth]{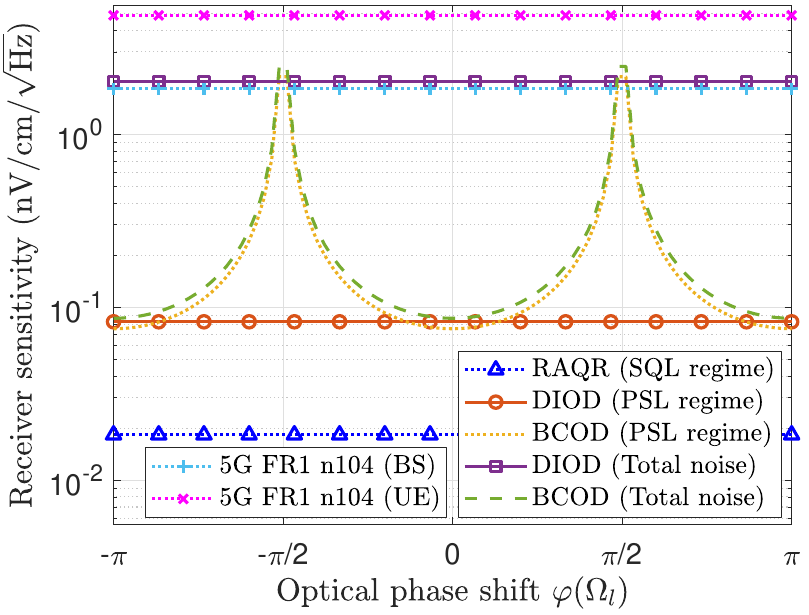}}
			\subfloat[]{
				\includegraphics[width=0.3\textwidth]{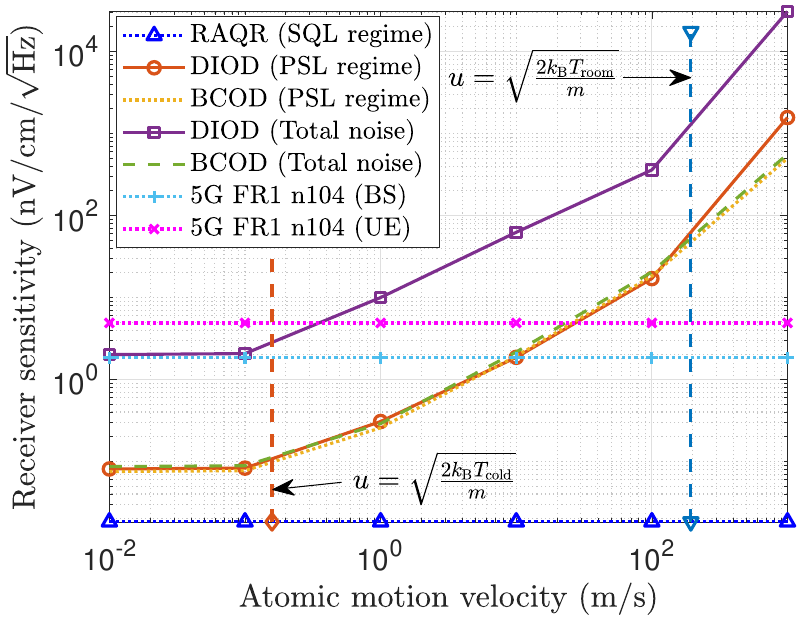}}
			\vspace{-0.5em}
			\caption{The SNR ratio and receiver sensitivity vs. (a)(d) environmental temperature, (b)(e) phase shift, and (c)(f) atomic velocity.}
			\vspace{-1.5em}
			\label{fig:sim_result_4}
		\end{figure*}

		\vspace{-1em}
		\subsection{Influence of Practical Impairments}
		\label{subsec:impairments}
		
		In this section, we investigate the performance of RAQRs influenced by several practical impairments, such as the environmental temperature, mismatch of optical phase shift, and the atomic motion velocity imposing Doppler broadening. This facilitates the understanding of performance mismatches between theory and practice, as well as guide future improvements of realistic experiments.

		\textbf{SNR ratio and receiver sensitivity vs. the environmental temperature:} 
		In realistic applications, the environmental temperature may dynamically change, which further influences the pressure inside the vapor cell according to the relationship of $P = 10^{9.717 - 3999 / T}$, as well as the atomic density via $N_0 = \frac{0.7217 P}{k_B T}$ \cite{rotunno2023investigating}. Intuitively, the rise of the environmental temperature increases both the cell's pressure and its atomic density. Based on these relationships, we plot the SNR ratio and receiver sensitivity with respect to the environmental temperature in Fig. \ref{fig:sim_result_4}(a),(d), respectively. 
		
		It can be observed from Fig. \ref{fig:sim_result_4}(a) that the SNR ratio (PSL regime) first increases to a maximum and then decreases when the environmental temperature varies from low to high. By contrast, the receiver sensitivity changes obeying the opposite trend, as seen in Fig. \ref{fig:sim_result_4}(d). Briefly, this phenomenon is caused by the competition between signal enhancement due to increased atomic density and signal attenuation due to the substantial absorption of the probe beam. We also note that the large cell's pressure at high environmental temperatures will lead to intensified atomic collisions that cause high-temperature decoherence and hence degrade the performance. In practical applications, it is critical to retain a stable environmental temperature for the vapor cell. Designing advanced temperature-controlled systems and/or alternatively resorting to advanced fabrications of vapor cells \cite{li2025wafer} may become a promising research direction.

		\textbf{SNR ratio and receiver sensitivity vs. the optical phase shift $\varphi ({\Omega _l})$:} 
		As observed from \eqref{eq:NarrowbandReceiveSNR}, the maximum of the received SNR is achieved when $\cos^2 {\varphi ({\Omega _l})} = 1$. This condition always holds for the DIOD scheme according to \eqref{eq:varphi_1}, while requiring fine-tuning $\phi_l$ of the local optical beam to guarantee $\varphi_2 (\Omega_l) = 0$, $\pm \pi$ according to \eqref{eq:varphi}. To investigate the mismatch of this optical phase shift, we present the SNR ratio and receiver sensitivity with respect to this mismatch in Fig. \ref{fig:sim_result_4}(b) and Fig. \ref{fig:sim_result_4}(e), respectively. 
		
		It can be seen in Fig. \ref{fig:sim_result_4}(b) that the SNR ratio of the BCOD scheme operating in the PSL regime gradually drops as $\varphi ({\Omega _l})$ moves away from $0, \pm \pi$, while the worst case occurs at $\pm \pi/2$. This also applies to the receiver sensitivity, which however exhibits the opposite trend, as portrayed in Fig. \ref{fig:sim_result_4}(e). 
		Furthermore, it is robust across a wide mismatch region, where the RAQRs exhibit their performance advantages over the classical 5G-BS (5G-UE), as observed from both sub-figures.

		\textbf{SNR ratio and receiver sensitivity vs. the atomic motion velocity:} 
		The thermally induced motion of atoms causes Doppler broadening, which broadens the optical resonance linewidth and significantly reduces both the SNR ratio and the receiver sensitivity. Higher atomic motion velocity further degrades the performance of RAQRs. This Doppler broadening effect is typically included in the relationship $\rho_{21}^{D} = \frac{1}{\sqrt{\pi} u} \int_{-\infty}^{\infty} \rho_{21} \left(\Delta_{p}^{\prime}, \Delta_{c}^{\prime}\right) e^{-\frac{v^{2}}{u^{2}}} dv$ \cite{holloway2017electric}, where $u = \sqrt{ \frac{2 k_{\text{B}} T}{m} }$ is the most likely atomic motion velocity at temperature $T$; $m = 2.206 \times 10^{-25}$ kg is the mass of a single Cs atom; $\Delta_{p}^{\prime}=\Delta_{p}-\frac{2 \pi}{\lambda_{p}} v$ and $\Delta_{c}^{\prime}=\Delta_{c}+\frac{2 \pi}{\lambda_{c}} v$ represent the modified detuning of the counter-propagating probe and coupling beams. In practice, the integration range can be selected as $[-au, au]$, where $a \ge 3$.
		
		To explicitly show this effect, we plot the SNR ratio and receiver sensitivity with respect to the atomic motion velocity in Fig. \ref{fig:sim_result_4}(c) and Fig. \ref{fig:sim_result_4}(f), respectively. It can be observed that the SNR ratio (in the PSL regime) and receiver sensitivity (in both the PSL and total noise regime) decreases and increases, respectively, as the atomic motion velocity varies from low to high. We also mark the most likely atomic motion velocity $u = \sqrt{ \frac{2 k_{\text{B}} T_{\text{room}}}{m} }$ at room temperature, where the RAQRs exhibit inferior performance compared to classical 5G-BS and 5G-UE. This deleterious effect of Doppler broadening can be substantially mitigated by reducing the atomic motion velocity, as observed from the most likely atomic motion velocity $u = \sqrt{ \frac{2 k_{\text{B}} T_{\text{cold}}}{m} }$ at an equivalent low temperature of $T_{\text{cold}} = 200$ \textmu K marked in both sub-figures. 
		Furthermore, we emphasize that optical cooling technologies, such as the magneto-optical trap (MOT) \cite{Tu2024Approaching}, may be employed for reducing the atomic kinetic energy to the \textmu K–mK level without requiring a cryogenic environment, where the thermal velocity variation almost disappears and hence the Doppler broadening can be substantially mitigated.

		\section{Discussions}
		\label{sec:Discussions}

		\textbf{Bandwidth adaptability:} 
		The proposed signal models are mainly valid for narrowband receptions, where the bandwidth of received RF signals should be within the $3$ dB bandwidth of RAQRs. As such, the steady-state of the quantum response can be effectively achieved, and thereby the waveform of the RF signal can be captured to yield a less distorted probe output within the atomic coherence time ($T_2$) of RAQRs. By contrast, the narrowband assumption becomes violated when the wireless fading channel is frequency-selective. For example, if the coherence time of the wireless fading channel is less than the atomic coherence time ($T_c \le T_2$), RAQRs would no longer reach the steady-state, and hence the steady-state solution of the Lindblad master equation becomes invalid. Instead, a time-dependent approach, addressing the differential equation of the Lindblad master equation, should be employed, where the Hamiltonian explicitly accounts for the time-varying RF amplitude corresponding to the channel fluctuations. This transient solution represents the interplay between the channel dynamics and the atomic coherence evolution. 
		Even so, the maximal supported bandwidth related to each Rydberg atomic transition is still restricted by the underlying physical properties. Given their narrowband but high-sensitivity characteristics, RAQRs may find their potential applications for small payloads or long-distance communication and sensing. These applications are becoming increasingly important in the context of future integrated space-air-ground-sea networks.

		\textbf{BBR modeling revisited:} 
		The BBR–induced noise may be modeled as an additive electric-field noise source, as adopted in \cite{Tu2024Approaching}, which obeys the conventional paradigm of classical RF receivers. This treatment potentially provides a practical engineering-oriented perspective for estimating the BBR–induced noise contribution. However, it may fail to accurately represent the dissipative quantum nature of the BBR coupling. Alternatively, BBR may be described in terms of radiation-induced transitions that affect both the decay rate and coherence time, thereby indirectly influencing the RAQR's performance. This treatment is supported by previous studies \cite{gallagher1994,Fancher2021Rydberg,beterov2009quasiclassical} and forms the basis of the modeling framework employed in this article. To the best of our knowledge, there is currently no rigorous evidence to support which treatment is more appropriate, or whether the two treatments are equivalent for RAQRs. Further investigation of the BBR-induced noise remains an important direction for future research, particularly toward establishing a rigorous connection between BBR-induced decoherence and classical additive-noise representations in RAQRs.

		\textbf{Experimental verifiability:}
		Although this treatise is primarily theoretical, the proposed models can be experimentally validated using a four-level superheterodyne scheme. By appropriately tuning the laser and RF parameters, and matching the parameters of this article to those employed in experiments, one can obtain the beat note waveform, input-output response curve, received SNR, and receiver sensitivity corresponding to different model parameters. These measurements can be directly compared to the theoretically predicted counterparts of this article. Such comparisons would provide a solid basis for validating the theoretical framework presented in this study. To arrive at a better consistency between theory and experiments, the above-mentioned experimental measurements may be performed block by block of the holistic RAQR-aided wireless receiver. Furthermore, the RF signal's bandwidth is limited within the $3$ dB bandwidth of RAQRs, so the RF signal's waveform can be captured within the response period of RAQRs. Indeed, we concur that there may exist discrepancies between the theory and experiments due to practical impairments, but our theoretical framework can characterize RAQRs by extracting their mathematical essence. Again, this is somewhat reminiscent of the role of Shannon's capacity, which reflects the essence of communications, despite its discrepancies compared to practical communication systems.

		\textbf{Practical challenges and potential solutions:}
		To fully harness the potential of RAQRs in real-world scenarios, several practical challenges should be addressed before either large-scale commercial implementations. Firstly, the requirement for narrow-linewidth, frequency-stabilized lasers and precise optical alignment increases system complexity, cost, size, weight, and power (C-SWaP). Chip-scale vapor cells, photonic integration, and compact laser modules offer promising routes towards miniaturization and robustness. Secondly, maintaining atomic coherence under realistic environmental conditions is difficult due to thermal drift, laser noise, etc. Active temperature stabilization, laser frequency locking, and coherent population trapping based stabilization may significantly improve system stability. Finally, for practical applications such as communication, radar, and field metrology, integration with compact optics, robust packaging, and automated calibration become essential. Continued progress in atomic device engineering, integrated photonics, and quantum-domain control is expected to gradually bridge the gap between theoretical demonstrations and real-world deployments of RAQRs.

		\section{Conclusions and Outlook}
		\label{sec:Conclusions}
		
		In this article, we have proposed an end-to-end RAQR-aided wireless receiver scheme by presenting its functional blocks and constructed a corresponding equivalent baseband signal model. Our model follows a realistic signal reception flow and takes into consideration the actual implementation of the RAQRs, which makes our model practical. The proposed scheme and model may be readily harnessed for system design and signal processing for future RAQR-aided wireless communications and sensing. We have further studied the DIOD and BCOD photodetection schemes, and have theoretically demonstrated that the BCOD scheme always outperforms the DIOD scheme. We have also compared the RAQRs to the classic RF receivers and shown the superiority of RAQRs in achieving a substantial received SNR gain of over $27$ dB and $40$ dB in the PSL and SQL regimes, respectively.

		Our results unveil the great potential of RAQRs and offer an easy-to-handle framework for future studies. Our signal model is expected to facilitate the following aspects
		\begin{enumerate}[label={\em {A\arabic*}}, leftmargin=1.7em, labelindent=0pt, itemindent = 0em]
			
			\item \label{itm:A1}  
			The current signal model is conceived for narrowband single-input single-output (SISO) systems, which serves as a basis to construct signal models for complex wireless systems, e.g., wideband and MIMO systems.
			
			\item \label{itm:A2}  
			Based on the proposed signal model, theoretical studies and system design are facilitated for RAQR-aided wireless systems. Our model also allows system comparisons in a convenient way.

			\item \label{itm:A3}  
			The proposed signal model facilitates the development of advanced signal processing approaches and optimization algorithms for realizing various upper-level objectives in the RAQR-aided wireless systems.  
		\end{enumerate}
		
		To exploit the potential of RAQRs, we foresee the following potent research directions for future RAQR-aided wireless communications and sensing. 
		\begin{enumerate}[label={\em {D\arabic*}}, leftmargin=1.7em, labelindent=0pt, itemindent = 0em]
			
			\item \label{itm:D1}  
			Wideband RAQRs: The current study is limited to specific discrete frequencies that are determined by the transitions of Rydberg states. The practical wireless communication and sensing applications prefer the wideband receiving and processing capabilities of RAQRs.
			
			\item \label{itm:D2}  
			RAQ-MIMO: The current implementations are essentially restricted to a single pair of laser beams to form a SISO system. The great potential of RAQRs is expected to be further exploited to form an RAQ-MIMO system. 
			
			\item \label{itm:D3}  
			Practical applications: The current study hinges on simplified assumptions and theoretical models. But the influence of practical impairments should be thoroughly investigated to explore the realistic boundaries and facilitate the practical deployments. 
		
		\end{enumerate}


		\appendices
		
		\section{The Coefficients of \eqref{eq:rho_21}} 
		\label{DensityMatrixCoeffs}
		\vspace{-1.8em}
		\begin{align}
			\label{eq:A1}
			&{A_1} =  - 2{\Delta _p}, \\
			\nonumber
			&{A_2} = 16{\Delta_c^2} \Delta_p +32\Delta_c {\Delta_p^2} + 16\Delta_l \Delta_c \Delta_p -2\Delta_c {\Omega_c^2} \\
			\label{eq:A2}
			&+16{\Delta_p^3} +16\Delta_l {\Delta_p^2} -2\Delta_p {\Omega_c^2} -2\Delta_l {\Omega_c^2}, \\
			\nonumber
			&{A_3} = -32{\Delta_c^4} \Delta_p -64{\Delta_c^3} \Delta_l \Delta_p -128{\Delta_c^3} {\Delta_p^2} +8{\Delta_c^3} {\Omega_c^2} \\
			\nonumber
			&-32{\Delta_c^2} {\Delta_l^2} \Delta_p -192{\Delta_c^2} \Delta_l {\Delta_p^2} +16{\Delta_c^2} \Delta_l {\Omega_c^2} -192{\Delta_c^2} {\Delta_p^3} \\
			\nonumber
			&+24{\Delta_c^2} \Delta_p {\Omega_c^2} -64\Delta_c {\Delta_l^2} {\Delta_p^2} +8\Delta_c {\Delta_l^2} {\Omega_c^2} -192\Delta_c \Delta_l {\Delta_p^3} \\
			\nonumber
			&+32\Delta_c \Delta_l \Delta_p {\Omega_c^2} -128\Delta_c {\Delta_p^4} +24\Delta_c {\Delta_p^2} {\Omega_c^2} -32{\Delta_l^2} {\Delta_p^3} \\
			\nonumber
			&+8{\Delta_l^2} \Delta_p {\Omega_c^2} -64\Delta_l {\Delta_p^4} +16\Delta_l {\Delta_p^2} {\Omega_c^2} \\
			\label{eq:A3}
			&-32{\Delta_p^5} +8{\Delta_p^3} {\Omega_c^2}, \\
			\label{eq:B1}
			&{B_1} = {\gamma _2}, \\
			\label{eq:B2}
			&{B_2} = 8 \gamma_2 \left( - {\Delta_c^2} -2 \Delta_c \Delta_p - \Delta_l \Delta_c - {\Delta_p^2} - \Delta_l \Delta_p \right), \\
			\nonumber
			&{B_3} = 16\gamma_2 {\Delta_c^4} +32\gamma_2 {\Delta_c^3} \Delta_l +64\gamma_2 {\Delta_c^3} \Delta_p +16\gamma_2 {\Delta_c^2} {\Delta_l^2} \\
			\nonumber
			&+96\gamma_2 {\Delta_c^2} \Delta_l \Delta_p +96\gamma_2 {\Delta_c^2} {\Delta_p^2} +32\gamma_2 \Delta_c {\Delta_l^2} \Delta_p \\
			\nonumber
			&+96\gamma_2 \Delta_c \Delta_l {\Delta_p^2} +64\gamma_2 \Delta_c {\Delta_p^3} +16\gamma_2 {\Delta_l^2} {\Delta_p^2} \\
			\label{eq:B3}
			&+32\gamma_2 \Delta_l {\Delta_p^3} +16\gamma_2 {\Delta_p^4}, \\
			\label{eq:C1}
			&{C_1} = 4\Delta _p^2 + 2\Omega _p^2 + \gamma _2^2, \\
			\nonumber
			&{C_2} = -32{\Delta_c^2} {\Delta_p^2} -16{\Delta_c^2} {\Omega_p^2} -8{\Delta_c^2} {\gamma_2^2} -64\Delta_c {\Delta_p^3} \\
			\nonumber
			&-32\Delta_l \Delta_c {\Delta_p^2} +8\Delta_c \Delta_p {\Omega_c^2} -32\Delta_c \Delta_p {\Omega_p^2} -16\Delta_c \Delta_p {\gamma_2^2} \\
			\nonumber
			&-16\Delta_l \Delta_c {\Omega_p^2} -8\Delta_l \Delta_c {\gamma_2^2} -32{\Delta_p^4} -32\Delta_l {\Delta_p^3} +8{\Delta_p^2} {\Omega_c^2} \\
			\nonumber
			&-16{\Delta_p^2} {\Omega_p^2} -8{\Delta_p^2} {\gamma_2^2} +8\Delta_l \Delta_p {\Omega_c^2} -16\Delta_l \Delta_p {\Omega_p^2} \\
			\label{eq:C2}
			&-8\Delta_l \Delta_p {\gamma_2^2} +2{\Omega_c^2} {\Omega_p^2} +2{\Omega_p^4}, \\
			\nonumber
			&{C_3} = 64{\Delta_c^4} {\Delta_p^2} +32{\Delta_c^4} {\Omega_p^2} +16{\Delta_c^4} {\gamma_2^2} +128{\Delta_c^3} \Delta_l {\Delta_p^2} \\
			\nonumber
			&+64{\Delta_c^3} \Delta_l {\Omega_p^2} +32{\Delta_c^3} \Delta_l {\gamma_2^2} +256{\Delta_c^3} {\Delta_p^3} -32{\Delta_c^3} \Delta_p {\Omega_c^2} \\
			\nonumber
			&+128{\Delta_c^3} \Delta_p {\Omega_p^2} +64{\Delta_c^3} \Delta_p {\gamma_2^2} +64{\Delta_c^2} {\Delta_l^2} {\Delta_p^2} +32{\Delta_c^2} {\Delta_l^2} {\Omega_p^2} \\
			\nonumber
			&+16{\Delta_c^2} {\Delta_l^2} {\gamma_2^2} +384{\Delta_c^2} \Delta_l {\Delta_p^3} -64{\Delta_c^2} \Delta_l \Delta_p {\Omega_c^2} \\
			\nonumber
			&+192{\Delta_c^2} \Delta_l \Delta_p {\Omega_p^2} +96{\Delta_c^2} \Delta_l \Delta_p {\gamma_2^2} +384{\Delta_c^2} {\Delta_p^4} \\
			\nonumber
			&-96{\Delta_c^2} {\Delta_p^2} {\Omega_c^2} +192{\Delta_c^2} {\Delta_p^2} {\Omega_p^2} +96{\Delta_c^2} {\Delta_p^2} {\gamma_2^2} +4{\Delta_c^2} {\Omega_c^4} \\
			\nonumber
			&+8{\Delta_c^2} {\Omega_c^2} {\Omega_p^2} +8{\Delta_c^2} {\Omega_p^4} +128\Delta_c {\Delta_l^2} {\Delta_p^3} -32\Delta_c {\Delta_l^2} \Delta_p {\Omega_c^2} \\
			\nonumber
			&+64\Delta_c {\Delta_l^2} \Delta_p {\Omega_p^2} +32\Delta_c {\Delta_l^2} \Delta_p {\gamma_2^2} +384\Delta_c \Delta_l {\Delta_p^4} \\
			\nonumber
			&-128\Delta_c \Delta_l {\Delta_p^2} {\Omega_c^2} +192\Delta_c \Delta_l {\Delta_p^2} {\Omega_p^2} +96\Delta_c \Delta_l {\Delta_p^2} {\gamma_2^2} \\
			\nonumber
			&+8\Delta_c \Delta_l {\Omega_c^4} +16\Delta_c \Delta_l {\Omega_c^2} {\Omega_p^2} +8\Delta_c \Delta_l {\Omega_p^4} +256\Delta_c {\Delta_p^5} \\
			\nonumber
			&-96\Delta_c {\Delta_p^3} {\Omega_c^2} +128\Delta_c {\Delta_p^3} {\Omega_p^2} +64\Delta_c {\Delta_p^3} {\gamma_2^2} +8\Delta_c \Delta_p {\Omega_c^4} \\
			\nonumber
			&+16\Delta_c \Delta_p {\Omega_c^2} {\Omega_p^2} +16\Delta_c \Delta_p {\Omega_p^4} +64{\Delta_l^2} {\Delta_p^4} -32{\Delta_l^2} {\Delta_p^2} {\Omega_c^2} \\
			\nonumber
			&+32{\Delta_l^2} {\Delta_p^2} {\Omega_p^2} +16{\Delta_l^2} {\Delta_p^2} {\gamma_2^2} +4{\Delta_l^2} {\Omega_c^4} +8{\Delta_l^2} {\Omega_c^2} {\Omega_p^2} \\
			\nonumber
			&+4{\Delta_l^2} {\Omega_p^4} +128\Delta_l {\Delta_p^5} -64\Delta_l {\Delta_p^3} {\Omega_c^2} +64\Delta_l {\Delta_p^3} {\Omega_p^2} \\
			\nonumber
			&+32\Delta_l {\Delta_p^3} {\gamma_2^2} +8\Delta_l \Delta_p {\Omega_c^4} +16\Delta_l \Delta_p {\Omega_c^2} {\Omega_p^2} +8\Delta_l \Delta_p {\Omega_p^4} \\
			\nonumber
			& +64{\Delta_p^6} -32{\Delta_p^4} {\Omega_c^2} +32{\Delta_p^4} {\Omega_p^2} +16{\Delta_p^4} {\gamma_2^2} +4{\Delta_p^2} {\Omega_c^4} \\
			\label{eq:C3}
			&+8{\Delta_p^2} {\Omega_c^2} {\Omega_p^2} +8{\Delta_p^2} {\Omega_p^4}. 
		\end{align}

		\section{The Proofs of \eqref{eq:PassbandRFplusLO} and \eqref{eq:Amp_PassbandRFplusLO}}
		\label{RFsuperpositionProof}

		The power of the superposition $z(t)$ can be obtained by its equivalent baseband signal $z_b(t) = {x_b}(t){e^{\jmath 2\pi {f_\delta }t}} + {y_b}(t)$ as 
		\begin{align}
			\nonumber
			\mathcal{P}_{z} &= z_b^2\left( t \right) 
			= \left[ {{x_b}(t){e^{\jmath 2\pi {f_\delta }t}} + {y_b}(t)} \right]{\left[ {{x_b}(t){e^{\jmath 2\pi {f_\delta }t}} + {y_b}(t)} \right]^*} \\
			\nonumber
			&= \mathcal{P}_{x} + \mathcal{P}_{y} + \sqrt{\mathcal{P}_{x} \mathcal{P}_{y}} \left[ {{e^{\jmath \left( {2\pi {f_\delta }t + {\theta _\delta }} \right)}} + {e^{ - \jmath \left( {2\pi {f_\delta }t + {\theta _\delta }} \right)}}} \right] \\
			\nonumber
			&= \mathcal{P}_{x} + \mathcal{P}_{y} + 2\sqrt{\mathcal{P}_{x} \mathcal{P}_{y}} \cos \left( {2\pi {f_\delta }t + {\theta _\delta }} \right). 
		\end{align}
		Therefore, based on the relationship of $\mathcal{P}_{z} = \frac{1}{2} c \epsilon_0 A_{e} \left| U_{z} \right|^{2}$, the amplitude $U_{z}$ is formulated as
		\begin{align}
			\nonumber
			{U_z} = \sqrt {\left| U_{x} \right|^{2} + \left| U_{y} \right|^{2} + 2{U_x}{U_y}\cos \left( {2\pi {f_\delta }t + {\theta _\delta }} \right)}.
		\end{align}
		We then write $z_b(t) = \sqrt{\mathcal{P}_{x}}{e^{\jmath {\theta _x}}}{e^{\jmath 2\pi {f_\delta }t}} + \sqrt{\mathcal{P}_{y}}{e^{\jmath {\theta _y}}}$ in the form of its real and imaginary parts as follows
		\begin{align}
			\nonumber
			{z_b}\left( t \right) 
			&= \sqrt{\mathcal{P}_{x}} \cos \left( {2\pi {f_\delta }t + {\theta _x}} \right) + \sqrt{\mathcal{P}_{y}} \cos \left( {{\theta _y}} \right) \\
			\nonumber
			&+ \jmath \left[ \sqrt{\mathcal{P}_{x}} \sin \left( {2\pi {f_\delta }t + {\theta _x}} \right) + \sqrt{\mathcal{P}_{y}} \sin \left( {{\theta _y}} \right) \right].
		\end{align}
		Let us denote the phase of $z_b(t)$ by $\theta_z$. We then obtain $\tan(\theta_z) \approx \tan(\theta_y)$ in \eqref{eq:phase}. Therein, $(1)$ holds using $A\sin(x) + B\sin(y) = (A+B)\sin(\frac{x+y}{2})\cos(\frac{x-y}{2}) + (A-B)\cos(\frac{x+y}{2})\sin(\frac{x-y}{2})$ and $A\cos(x) + B\cos(y) = (A+B)\cos(\frac{x+y}{2})\cos(\frac{x-y}{2}) - (A-B)\sin(\frac{x+y}{2})\sin(\frac{x-y}{2})$, respectively. The approximation $(2)$ holds relying on $\sqrt{\mathcal{P}_{y}} \gg \sqrt{\mathcal{P}_{x}}$.
		\begin{figure*}
			\begin{align}
				\nonumber
				\tan ({\theta _z}) 
				&\overset{(1)}{=} \frac{{\left( {\sqrt{\mathcal{P}_{x}} + \sqrt{\mathcal{P}_{y}}} \right)\sin \left( {\frac{{2\pi {f_\delta }t + {\theta _x} + {\theta _y}}}{2}} \right)\cos \left( {\frac{{2\pi {f_\delta }t + {\theta _x} - {\theta _y}}}{2}} \right) + \left( {\sqrt{\mathcal{P}_{x}} - \sqrt{\mathcal{P}_{y}}} \right)\cos \left( {\frac{{2\pi {f_\delta }t + {\theta _x} + {\theta _y}}}{2}} \right)\sin \left( {\frac{{2\pi {f_\delta }t + {\theta _x} - {\theta _y}}}{2}} \right)}}{{\left( {\sqrt{\mathcal{P}_{x}} + \sqrt{\mathcal{P}_{y}}} \right)\cos \left( {\frac{{2\pi {f_\delta }t + {\theta _x} + {\theta _y}}}{2}} \right)\cos \left( {\frac{{2\pi {f_\delta }t + {\theta _x} - {\theta _y}}}{2}} \right) - \left( {\sqrt{\mathcal{P}_{x}} - \sqrt{\mathcal{P}_{y}}} \right)\sin \left( {\frac{{2\pi {f_\delta }t + {\theta _x} + {\theta _y}}}{2}} \right)\sin \left( {\frac{{2\pi {f_\delta }t + {\theta _x} - {\theta _y}}}{2}} \right)}} \\
				\label{eq:phase}
				&\overset{(2)}{\approx} \frac{{\sin \left( {\frac{{2\pi {f_\delta }t + {\theta _x} + {\theta _y}}}{2}} \right)\cos \left( {\frac{{2\pi {f_\delta }t + {\theta _x} - {\theta _y}}}{2}} \right) - \cos \left( {\frac{{2\pi {f_\delta }t + {\theta _x} + {\theta _y}}}{2}} \right)\sin \left( {\frac{{2\pi {f_\delta }t + {\theta _x} - {\theta _y}}}{2}} \right)}}{{\cos \left( {\frac{{2\pi {f_\delta }t + {\theta _x} + {\theta _y}}}{2}} \right)\cos \left( {\frac{{2\pi {f_\delta }t + {\theta _x} - {\theta _y}}}{2}} \right) + \sin \left( {\frac{{2\pi {f_\delta }t + {\theta _x} + {\theta _y}}}{2}} \right)\sin \left( {\frac{{2\pi {f_\delta }t + {\theta _x} - {\theta _y}}}{2}} \right)}}
				= \frac{{\sin \left( {{\theta _y}} \right)}}{{\cos \left( {{\theta _y}} \right)}} = \tan ({\theta _y}).
			\end{align}
			\hrulefill
			\vspace{-0.2cm}
		\end{figure*}
		Further applying the Taylor series expansion to the above $U_z$, we have \eqref{eq:Amp_PassbandRFplusLO}. The proofs are completed.

		\vspace{-0.2cm}
		\section{The Proof of \eqref{eq:PhotodetectorOutputBCOD}}
		\label{BCODOutputSignalExpressionProof}
		
		We first derive the expression of $V^{\text{(B)}}(t) = \sqrt{G} I_{\text{ph}}^{\text{(B)}} (t)$ as
		\begin{align}
			\nonumber
			V^{\text{(B)}}(t) 
			&= \sqrt{G} \alpha \Big[ {P_{lb}}\left( t \right){P_b^*}\left( {{\Omega _{{\rm{RF}}}},t} \right) + P_{lb}^*\left( t \right) P_b \left( {{\Omega _{{\rm{RF}}}},t} \right) \Big] \\
			\nonumber
			&= 2 \sqrt{G} \alpha \sqrt{\mathcal{P}_{l} \mathcal{P}_{1} (\Omega_{\rm{RF}})} \cos \left( {{\phi _l} - {\phi _p}({\Omega _{{\rm{RF}}}})} \right).
		\end{align}
		We exploit $\Omega_{l} \gg \Omega_{x}$ and express $P_{\text{mix}} (\Omega_{\text{RF}}) \triangleq \sqrt{2 \mathcal{P}_{1} (\Omega_{\rm{RF}})} \cos \left( {{\phi _l} - {\phi _p}({\Omega _{{\rm{RF}}}})} \right)$ in the vicinity of a given point $\Omega_{\text{RF}} = \Omega_l$ using the Taylor series expansion. Upon retaining the first-order term and ignoring the high-order terms, we have $P_{\text{mix}}(\Omega_{\text{RF}}) \approx P_{\text{mix}}(\Omega_{l}) + P'_{\text{mix}}(\Omega_{l}) \left( \Omega_{\text{RF}} - \Omega_{l} \right)$. Specifically, the first-order derivative is derived as  
		\begin{align}
			\nonumber
			&P'_{\text{mix}} (\Omega_{\text{RF}}) \big|_{ \Omega_{\text{RF}} = \Omega_{l} }
			= {\left[ \sqrt{2 \mathcal{P}_{1} (\Omega_{\rm{RF}})} \cos \left( {{\phi _l} - {\phi _p}({\Omega _{{\rm{RF}}}})} \right) \right]^\prime } \Big|_{ \Omega_{\text{RF}} = \Omega_{l} } \\
			\nonumber
			&= \left( { - \tfrac{{\pi d}}{{{\lambda _p}}}} \right) \sqrt{2 \mathcal{P}_{1} (\Omega_{l})} \Big[ \cos \left( {{\phi _l} - {\phi _p}({\Omega _l})} \right){\mathscr I}\{ {\chi ^\prime }({\Omega _l})\} \Big. \\ 
			\nonumber
			&\Big. \qquad \qquad \qquad \qquad \qquad \quad - \sin \left( {{\phi _l} - {\phi _p}({\Omega _l})} \right) {\mathscr R}\{ {\chi ^\prime }({\Omega _l})\} \Big].
		\end{align}
		Upon defining ${{\mathscr I}\{ {\chi' }({\Omega _l})\} } = A \cos{\psi_p({\Omega _l})}$ and ${{\mathscr R}\{ {\chi' }({\Omega _l})\} } = A \sin{\psi_p({\Omega _l})}$, where $A \triangleq \sqrt {{{\left[ {{\mathscr I}\{ \chi'({\Omega _l})\} } \right]}^2} + {{\left[ {{\mathscr R}\{ \chi'({\Omega _l})\} } \right]}^2}}$ and $\psi_p({\Omega _l}) = \arccos \frac{ {{\mathscr I}\{ \chi '({\Omega _l})\} } }{ A }$, we have 
		\begin{align}
			\nonumber
			P'_{\text{mix}} (\Omega_{\text{RF}}) \big|_{ \Omega_{\text{RF}} = \Omega_{l} }
			&= - \tfrac{{\pi d}}{{{\lambda _p}}} \sqrt{2 \mathcal{P}_{1} (\Omega_{l})} \\
			\nonumber
			&\times A \cos \left( {\phi _l} - {\phi _p}({\Omega _l}) + \psi_p({\Omega _l}) \right).
		\end{align}
		Further exploiting the relationship $\Omega_x = \frac{ \mu_{34} }{ \hbar } U_x$, and defining $\kappa_2 (\Omega_l) \triangleq \frac{\pi d \mu_{34}}{\hbar \lambda_p} A$ and ${\varphi_2({\Omega _l})} \triangleq {\phi _l} - {\phi _p}({\Omega _l}) + {\psi _p}({\Omega _l})$, we have $V^{\text{(B)}}(t)$ approximated to 
		\begin{align}
			\nonumber
			V^{\text{(B)}}(t) 
			&\approx \sqrt{G} \alpha \sqrt{2\mathcal{P}_{l}} P_{\text{mix}}(\Omega_{l}) \\
			\nonumber
			&\;\;\; + \sqrt{G} \alpha \sqrt{2\mathcal{P}_{l}} P'_{\text{mix}}(\Omega_{l}) \left( \Omega_{\text{RF}} - \Omega_{l} \right) \\
			\nonumber
			&\approx 2\sqrt{G} \alpha \sqrt{\mathcal{P}_{l} \mathcal{P}_{1} (\Omega_{l})} 
			\Big[ \cos \left( {\phi _l} - {\phi _p}({\Omega _l}) \right) - \kappa_2({\Omega _l}) \Big. \\
			\nonumber
			&\Big. \qquad \times \cos {\varphi_2({\Omega _l})} U_x \cos \left( {2\pi {f_\delta }t + {\theta _\delta }} \right) \Big], 
		\end{align}
		which completes the proof.

\bibliographystyle{IEEEtran}
\bibliography{IEEEabrv,references}

\end{document}